\documentclass[namedreferences]{solarphysics}
\usepackage[optionalrh,solaromanenum]{spr-sola-addons} % For Solar Physics 
\usepackage[]{hyperref}
\hypersetup{
    colorlinks = true,
    allcolors = {blue},
}
\usepackage{graphicx}                    % For eps figures, newer & more powerfull
\usepackage{bm}
\usepackage{color}                       % For color text: \color command
\usepackage{breakurl}                         % For breaking URLs easily trough lines
\begin{document}

\begin{article}

\begin{opening}

\title{Numerical Simulations of Torsional Alfv\'{e}n Waves in Axisymmetric Solar Magnetic Flux Tubes}

%%%%%%%%%%%%%%%%%%%%%%%%%%%%%%%%%%%%%%%%%%%%%%%%%%%
%% Authors Names
%
 \author[addressref={1},corref,email={E-mail: dwojcik@kft.umcs.lublin.pl}]{\inits{D.}\fnm{D.}\lnm{W\'{o}jcik}\orcid{0000-0002-4200-3432}}
 \author[addressref={1},corref,email={}]{\inits{}\fnm{K.}\lnm{Murawski}}
 \author[addressref={2,3},corref,email={}]{\inits{}\fnm{Z.E.}\lnm{Musielak}}
 \author[addressref={1},corref,email={}]{\inits{}\fnm{P.}\lnm{Konkol}}
 \author[addressref={4},corref,email={}]{\inits{}\fnm{A.}\lnm{Mignone}}
%%%%%%%%%%%%%%%%%%%%%%%%%%%%%%%%%%%%%%%%%%%%%%%%%%%
%% Runningheads
%
\runningauthor{W\'{o}jcik et al.}
\runningtitle{Driven Alfv\'{e}n waves}

%%%%%%%%%%%%%%%%%%%%%%%%%%%%%%%%%%%%%%%%%%%%%%%%%%%
%% Affilations 
%% id shold be the same with \author addressref value.
\address[id={1}]{Group of Astrophysics, Faculty of Mathematics, Physics and Informatics, UMCS, ul. 
Radziszewskiego 20-031 Lublin, Poland}
\address[id={2}]{Department of Physics, University of Texas at Arlington, Arlington, TX 76019, USA}
\address[id={3}]{Kiepenheuer-Institut f\"{u}r Sonnenphysik Sch\"{o}neckstr. 6, 79104 Freiburg, Germany}
\address[id={4}]{Dipartimento di Fisica Generale Facolt\'a di Scienze M.F.N., 
Universit\'a degli Studi di Torino, 
%Via Pietro Giuria, 1
10125 Torino, Italy}

%%%%%%%%%%%%%%%%%%%%%%%%%%%%%%%%%%%%%%%%%%%%%%%%%%%
%%% Abstract 
\begin{abstract}

We investigate numerically Alfv\'{e}n waves propagating along an axisymmetric and non-isothermal solar flux tube embedded in the solar atmosphere. 
The tube magnetic field is current-free and diverges with height, and the waves are excited by a periodic driver along the tube magnetic field lines. 
The main results are that the two wave variables, the velocity and magnetic field perturbations in the azimuthal direction, behave differently as a result of gradients of physical parameters along the tube. 
To explain these differences in the wave behavior, the time evolution of the wave variables and the resulting cutoff period for each wave variable are calculated, 
and used to determine regions in the solar chromosphere where strong wave reflection may occur.  
\end{abstract}

%%%%%%%%%%%%%%%%%%%%%%%%%%%%%%%%%%%%%%%%%%%%%%%%%%%
%% Keywords
%

\end{opening}
%-------------------------------------------------

%%%%%%%%%%%%%%%%%%%%%%%%%%%%%%%%%%%%%%%%%%%%%%%%%%%
%% Sections
%
 \section{Introduction}%\label{s:?} 
Observations by several recently launched spacecrafts have
revealed the ubiquitous presence of oscillations in the solar
atmosphere, which can be interpreted as magnetohydrodynamics (MHD) waves \citep[\textit{e.g.}][]
{Nakariakov2005} or specifically as Alfv\'en waves whose signatures were observed in prominences, 
spicules and X-ray jets by \citet{Okamoto2007}, \citet{De_Pontieu2007} and \citet{Cirtain2007},
respectively.  Moreover, observational evidence for the existence of torsional Alfv\'en waves in the 
solar atmosphere was given by \citet{Jess2009}, see however \citet{Dwivedi2010} and 
\citet{Okamoto2011} who reported the presence of propagating Alfv\'en waves in solar spicules. 
Alfv\'en waves have been of a particular interest because they can carry energy and momentum along 
magnetic field lines to the solar corona, where the wave energy can heat the corona and the wave 
momentum may accelerate the solar wind.

There is a large body of literature devoted to Alfv\'en waves and their propagation in the solar atmosphere 
\citep{Zhugzhda1982,Hollweg1982,An1989,Hollweg1990,Musielak1992,MusielakMoore1995,KudohShibata1999,
HollwegIsenberg2007,Musielak2007,Murawski2010,Routh2010,Webb2012,Chmielewski2013,Murawski2014,Perera2015}. 
There are two main problems considered in these papers, namely, 
the propagation conditions for Alfv\'en waves and the dissipation of energy and momentum carried by these waves. 
The first problem involves the concept of cutoff frequency whose existence is caused by the presence of 
strong gradients of physical parameters in the solar atmosphere, and it has been explored for linear 
Alfv\'en waves by \citet{Murawski2010} and \citet{Perera2015}, and for torsional Alfv\'en waves in 
thin magnetic flux tubes by \citet{Musielak2007} and \citet{Routh2010}.  The second problem deals with 
the coronal heating \citep[]{Suzuki2005} 
and it involves different mechanisms of Alfv\'en wave dissipation, like phase-mixing \citep[]{Ofman1995} 
or nonlinear mode coupling \citep{Ulmschneider1991}, and wave momentum deposition \citep[]{Dwivedi2006,Chmielewski2013,Chmielewski2014}; 
however, a realistic modeling of both Alfv\'en wave propagation and dissipation is difficult to perform \citep[]{Banerjee1998,OShea2005,Bemporad2012,Chmielewski2013,Jelnek2015}.

\citet{Murawski2010} considered impulsively generated Alfv\'{e}n waves in the one-dimensional 
solar atmosphere with a smoothed step-wise temperature profile and a vertical magnetic field. 
\citet{Perera2015} studied analytically and numerically the case of periodically driven Alfv\'{e}n 
waves and their propagation in an isothermal solar atmosphere.  \citet{Musielak2007} and 
\citet{Routh2010} investigated torsional Alfv\'en waves propagating in thin magnetic flux
tubes embedded in the isothermal and non-isothermal solar atmosphere, respectively.  The main 
aim of this paper is to extend the work of \citet{Murawski2010} and \citet{Perera2015} to an 
axi-symmetric solar magnetic flux tube embedded in the solar atmosphere with a realistic 
temperature profile of \citet{Avrett2008} and curved magnetic field lines \citep{Low1985}, 
perform numerical simulations of the propagation of torsional Alfv\'{e}n waves, and to compare
the obtained numerical results to the analytical ones obtained by \citet{Routh2010}.
 
This paper is organized as follows.  The numerical model of the solar atmosphere is described
in Section 2.  The numerical and analytical results are presented and discussed in Sections 3
and 4, respectively.  Our discussion of the analytical and numerical result is given in Section 
5, and a short summary of the results and some concluding remarks are presented in Section 6.
\section{Numerical Model}
We consider a magnetically structured and gravitationally stratified
solar atmosphere, which is described by the following set of ideal
MHD equations:
\begin{eqnarray} 
\frac{\partial \varrho}{\partial t} + \nabla \cdot (\varrho \bm{V}) = 0 \,,\\ 
\varrho \frac{\partial \bm{V}}{\partial t}
+ \varrho (\bm{V} \cdot \nabla) \bm{V} &=
- \nabla p + \frac{1}{\mu} (\nabla \times \bm{B}) \times \bm{B}
+ \varrho \bm{g} \,, \\
\frac{\partial \bm{B}}{\partial t} &=
\nabla \times (\bm{V} \times \bm{B}) \,,
\qquad \nabla \cdot \bm{B} = 0 \,, \\
\frac{\partial p}{\partial t} + \bm{V} \cdot \nabla p &=
-\gamma p \nabla \cdot \bm{V} \,, \qquad
p = \frac{k_{\rm B}}{m} \varrho T \,,
\label{eq:mhd}     % No ``\\'' for the last equation
\end{eqnarray}
where $\varrho$ is mass density, $p$ is gas pressure, $\bm{V}$ is the
plasma velocity, $\bm{B}$ the magnetic field, and $\bm{g}=(0,-g,0)$
is the gravitational acceleration of magnitude $274~{\rm m}~{\rm s}^{-2}$.
The symbol $T$ denotes temperature, $m$ is particle mass, specified
by a mean molecular weight of $1.24$, $k_{\rm B}$ is the Boltzmann's
constant, $\gamma = 1.4$ is the adiabatic index, and $\mu$ the magnetic
permeability of plasma.

\begin{figure} 
      \vspace{5cm}
   \centerline{\includegraphics[width=5cm, height=4.5cm]{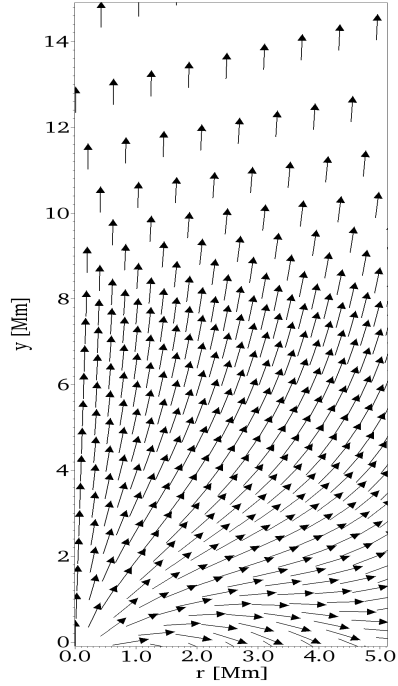}
              }
\caption{
Magnetic field vectors. %for $a=-0.75~{\rm Mm}$ in Eq.~(\ref{eq:equil:mag})
%\vspace{1.5cm}
}
\label{fig:equilibrium}
   \end{figure}

We consider the axisymmetric case in which all plasma variables are
invariant in the azimuthal direction $\theta$, but the $\theta$-components
of the perturbed magnetic field ${B_{\theta}}$ and velocity ${V_{\theta}}$ are not 
identically zero. We assume that the equilibrium magnetic field is current-free,
$\nabla \times \bm{B}= {\bf 0}$, and its components along the radial ($r$), azimuthal 
($\theta$), and vertical ($y$), directions are specified by
the following expressions: %(Low 1985)
\begin{eqnarray} 
B_{\rm{e}r} &=& \frac{3Sr(y-a)}{\left( r^2+(y-a)^2 \right)^\frac{5}{2}} \,, \\
B_{\rm{e} \theta} &=& 0 \,, \\
B_{\rm{e} y} &=& \frac{S\left(r^2-2(y-a)^2\right)}
{\left(r^2+(y-a)^2\right)^\frac{5}{2}} + B_{\rm{e}_0} \,,
\label{eq:equil:mag}
\end{eqnarray} 
where $a$ and $S$ are free parameters which denote the vertical location of 
the magnetic moment and the vertical magnetic field strength, respectively. 
Equations (5)--(7) comprise a special (potential and axisymmetric) 
case of the three-dimensional model which was developed by \citet{Low1985}. 
We set $a=-0.75~{\rm Mm}$ choose $S$ such that at $r=0~{\rm Mm}$ and 
$y_r=6.0~{\rm Mm}$ the magnetic field is $B_{\rm{e}}=B_{\rm{e}y}=9.5 ~{\rm Gs}$. Here, $y=y_r$ denotes the reference level. The magnetic 
field vectors resulting from these equations are drawn in Figure~\ref{fig:equilibrium}.
The magnetic field lines are curved and the curvature decays with height $y$,
it grows with the radial distance $r$, and the magnitude of ${\bf {B_{\rm{e}}}}$
decays with $y$ and $r$.

As the equilibrium magnetic field, specified by Equations (5)--(7) is current-free, 
the equilibrium gas pressure and mass density are hydrostatic and they are given
as \citep[\textit{e.g.}][]{Murawski2015}:
\begin{eqnarray} 
p_{\rm h}(y) &= p_0 \exp
\left[
-\int_{y_{\rm r}}^{y}
\frac{{\rm d}y'}{\Lambda(y')}
\right] \,,\hspace{3mm}
\varrho_{\rm h}(y)=-\frac{1}{g} \frac{\partial p_{\rm h}}{\partial y} \,,
\label{equalibrum_p}
\end{eqnarray} 
where $p_0=0.01~{\rm Pa}$
is the gas pressure evaluated at $y=y_{\rm r}$,
\begin{equation}
\Lambda(y) = \frac{k_{\rm B}T_{\rm h}(y)}{mg}
\end{equation}
is the pressure scale-height, and the symbol $T_{\rm h}(y)$ denotes
the hydrostatic temperature profile specified in the
model of \citet{Avrett2008}.

In the model, the solar photosphere occupies the region $0<y<0.5~{\rm Mm}$, 
the chromosphere resides in $0.5~{\rm Mm}<y<2.1~{\rm Mm}$, and the solar corona 
is represented by higher atmospheric layers, which start from the transition 
region that is located at $y \approx 2.1~{\rm Mm}$. Note that in equilibrium 
the gas pressure and mass density do not vary with the radial distance $r$, 
while they fall off with the height $y$ (not shown).

\begin{figure} 
   \centerline{\includegraphics[width=0.5\textwidth,clip=]{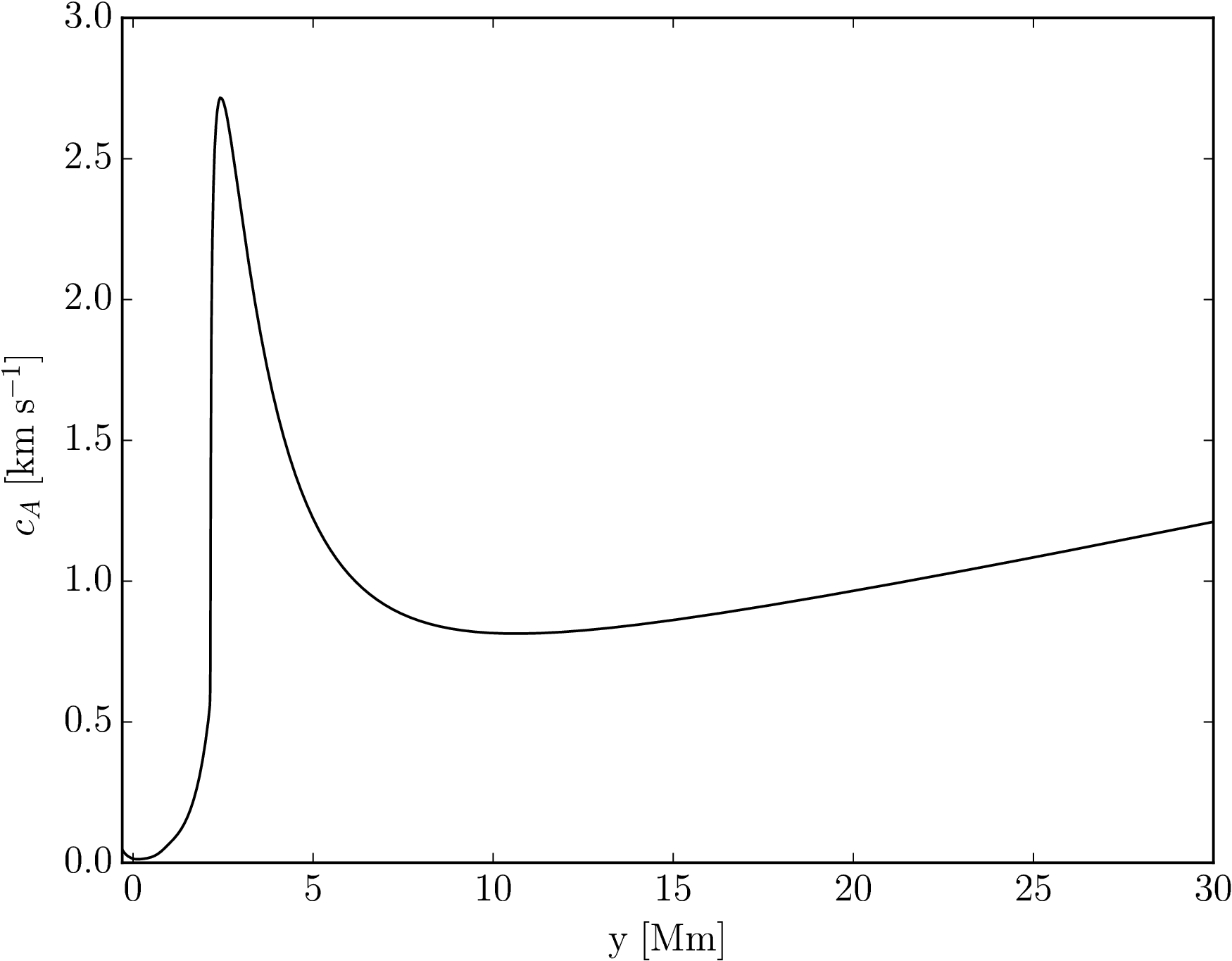}
              }
\caption{
Vertical profile of Alfv\'{e}n speed $c_{\rm A}$ for $r=0~{\rm Mm}$ 
at the equilibrium conditions of Equations. (5)-(8).}
\label{fig:cA}
   \end{figure}

Figure~\ref{fig:cA} shows the vertical profile of the Alfv\'{e}n speed 
$c_{\rm A}=B_e/\sqrt{\mu \varrho_h}$ along the central axis, 
$r=0~{\rm Mm}$, with $c_{\rm A}=12~{\rm km}~{\rm s^{-1}}$ at $y=0~{\rm Mm}$.
Higher up $c_{\rm A}$ grows with $y$, reaching
maximal value of $2715~{\rm km}~{\rm s^{-1}}$  at $y=2.44~{\rm Mm}$, and
then $c_{\rm A}$ falls off with height.  This characteristic shape of the
Alfv\'{e}n velocity profile will have a prominent effect on the wave behavior
in our model as shown and discussed in Section 3.

\section{Numerical Results}
Numerical simulations are performed using the PLUTO code, 
which is based on a finite-volume/finite-difference method 
\citep[]{Mignone2007,Mignone2011}, designed to solve
a system of conservation laws on structured meshes.
For all considered cases we set the simulation box as
0 to $5.12~{\rm Mm}$ in the $r$ direction and $-0.1~{\rm Mm}$ to $30~{\rm Mm}$ in the $y$ direction.
We impose the boundary conditions by fixing at the top and bottom of
the simulation region all plasma quantities to their equilibrium values,
while at the right boundary, outflow boundary conditions are implemented.
These outflow boundary conditions do no lead to any incoming signal reflection as we use the background magnetic field splitting \citep[]{Mignone2007}. 
At $r=0~{\rm Mm}$ axisymmetrical boundaries are implemented. 
Along the $r$-direction this simulation box is divided into $1024$ cells
and along the $y$-direction into $1536$ cells for $-0.1<y<7.58~{\rm Mm}$
and into $512$ cells in the range of $7.58<y<30~{\rm Mm}$.

For our problem the Courant-Friedrichs-Lewy number is set to $0.3$ and the
Harten-Lax-van Leer Discontinuities (HLLD) approximate Riemann solver\\ \citep[]{Toro2009} is adopted.
At the bottom boundary the periodic driver is additionally set as
\begin{equation}
V_\theta(r,t) = \frac{r}{w}   A_V 
\exp \left( -\frac{r^2}{w^2} \right) \sin\left(\frac{2\pi}{P_d}t\right) \,,
\end{equation}
where $P_{\rm d}$ is the period of the driver, $A_V$ is the
amplitude of the driver and $w$ is its spatial width.
We set and hold fixed $w=100~{\rm km}$ and $A_V=5~{\rm km}~{\rm s}^{-1}$, while allowing
$P_{\rm d}$ to vary.

\begin{figure} 
                                % includes the two top panels 
   \vspace{2cm}
   \centerline{
               \includegraphics[width=0.48\textwidth]{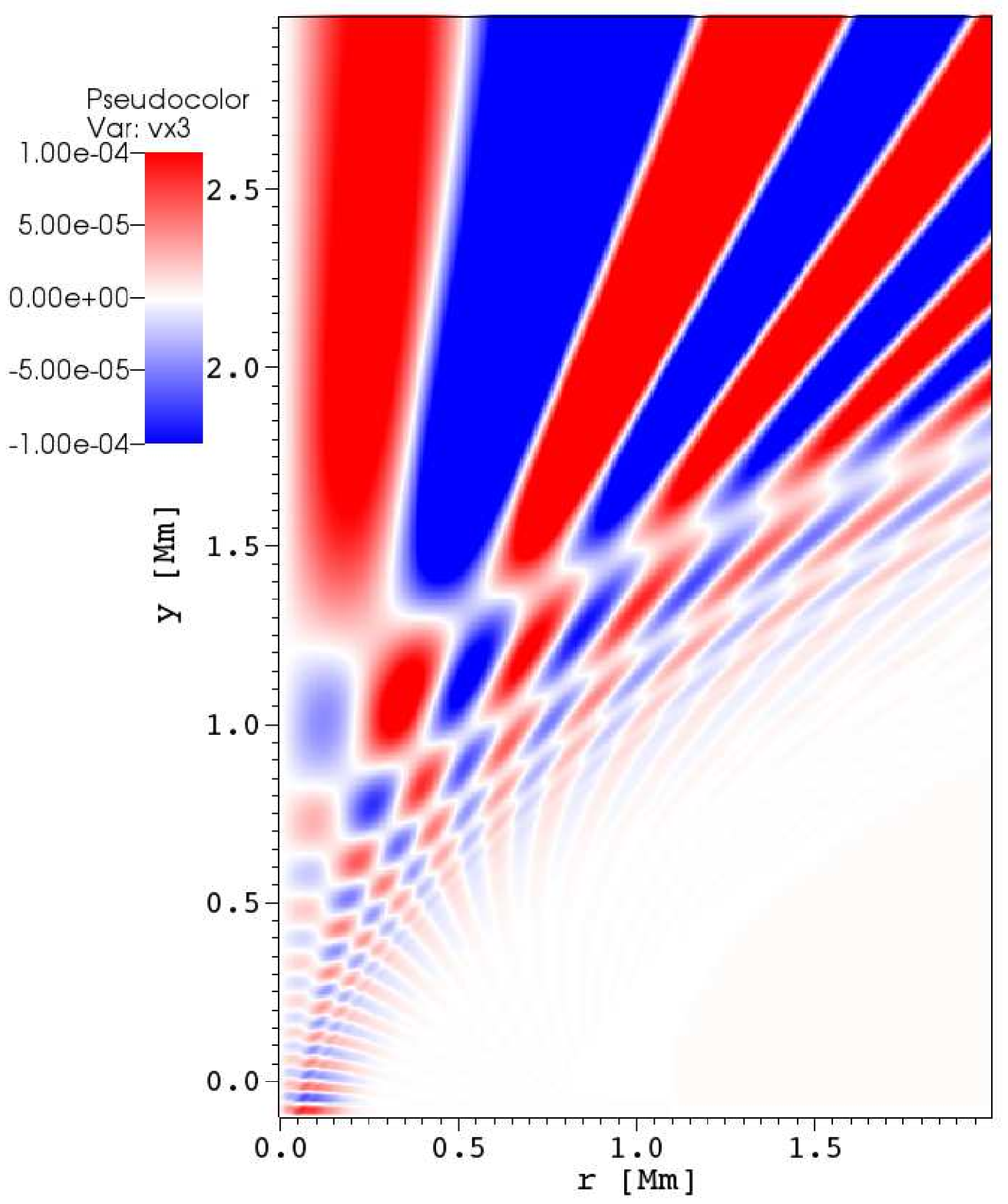}
               \hspace*{0.01\textwidth}
               \includegraphics[width=0.48\textwidth]{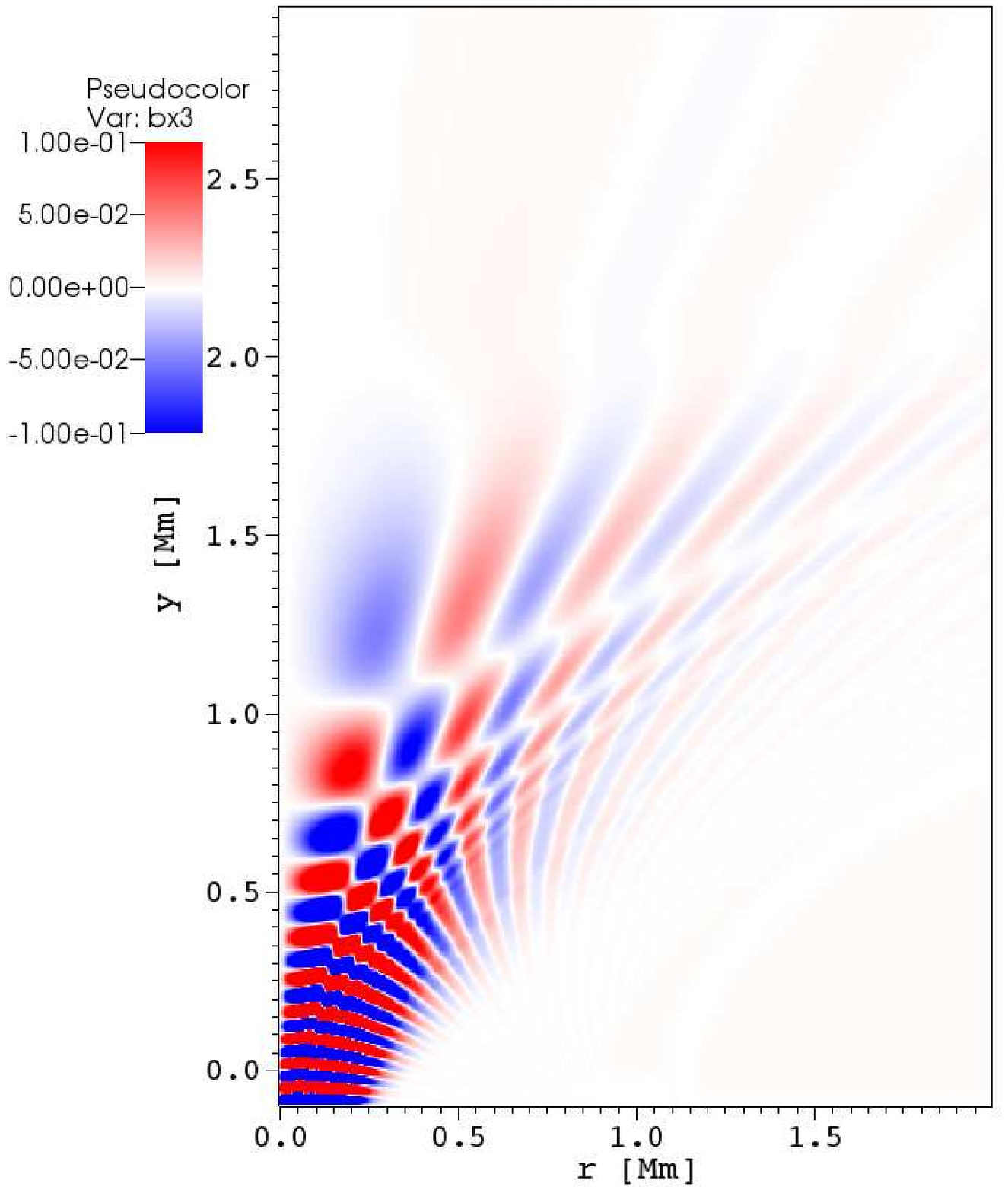}
              }        
\caption{Spatial profile of $V_\theta$ (left) and $B_\theta$ (right) for $P_d=35~\rm s$ at $t$=340~s.}
\label{fig:spat_prof_v}
   \end{figure}

Figure~\ref{fig:spat_prof_v} illustrates the $r$-$y$ spatial profiles of
$V_\theta$ (left) and $B_\theta$ (right) for $P_{\rm d}=35~{\rm s}$ 
which corresponds to an effective amplitude of the driver of about
$0.2~{\rm km}~{\rm s^{-1}}$.
It is seen in $V_\theta(r,y)$ that this driver excites Alfv\'{e}n waves which penetrate
the chromosphere, and at $t \approx 70~{\rm s}$ reach the transition region and the solar corona.
As a result of the Alfv\'{e}n speed profile, displayed in Figure~\ref{fig:cA}, the Alfv\'{e}n waves accelerate up to the altitude 
where $c_{\rm A}$ attains its maximal value, and higher up the Alfv\'{e}n waves decelerate.

\begin{figure}
                                % includes the two top panels 
   \centerline{\hspace*{0.015\textwidth}
               \includegraphics[width=0.5\textwidth,clip=]{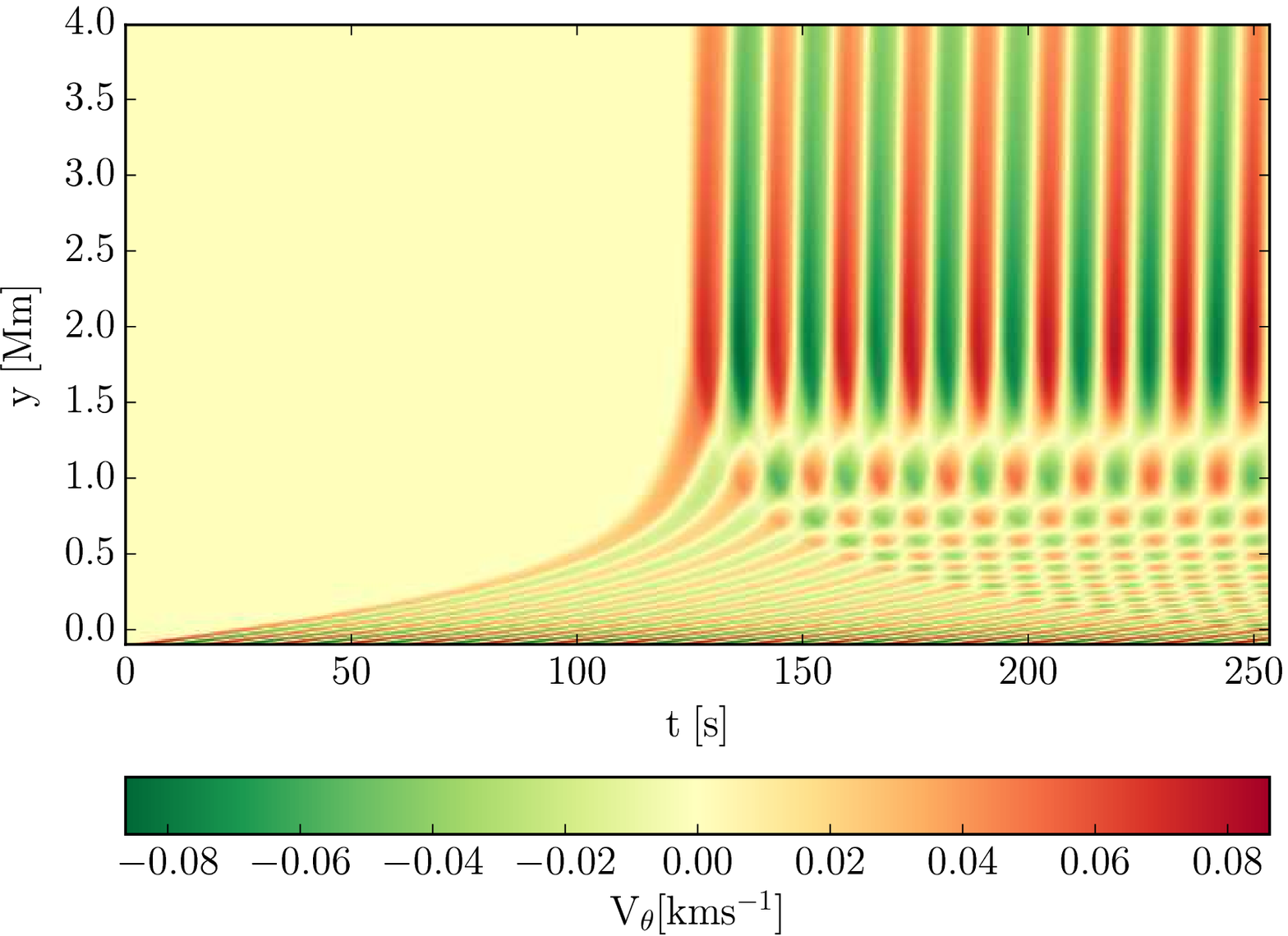}
               \hspace*{0.01\textwidth}
               \includegraphics[width=0.5\textwidth,clip=]{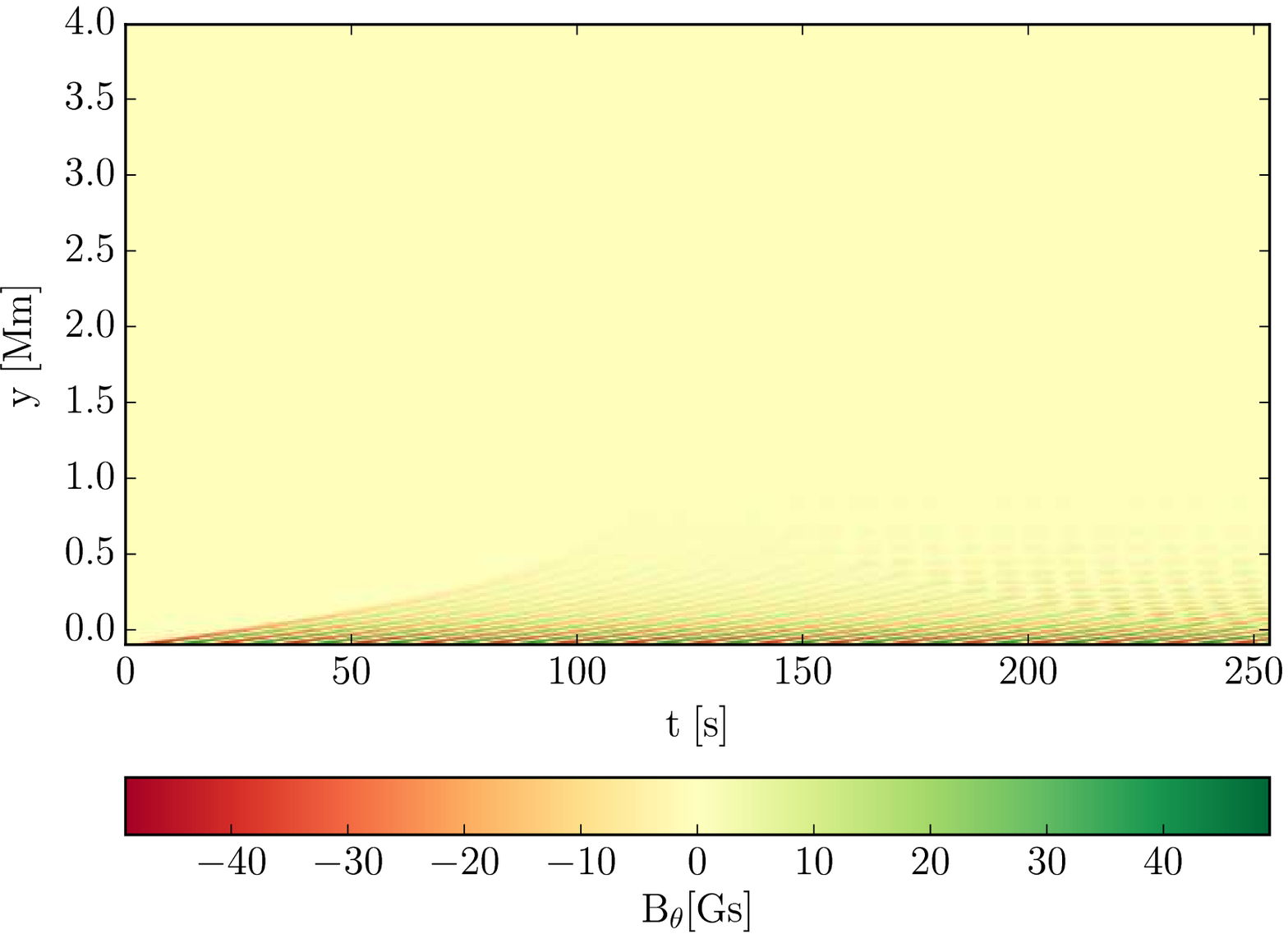}
              }        
\caption{
Vertical profiles of $V_\theta$ (left) and $B_\theta$ (right) for $r=0.1~{\rm Mm}$ \textit{vs} time for $P_d=35~\rm s$.}
\label{fig:vert_prof_v}
   \end{figure}

Note that as a consequence of the divergence of the magnetic field with height (Figure~\ref{fig:equilibrium}), 
the wave front spreads with height and magnetic shells develop in time \citep[]{Murawski2015shells} and a standing wave pattern is present in the low photosphere, below $y\approx0.5~{\rm Mm}$. 
A qualitatively similar scenario can be seen in the low atmospheric layers in profiles of $B_\theta(r,y)$ (Figure~\ref{fig:spat_prof_v}, right).
However, the perturbations in $B_\theta$ decay with altitude, remaining
largest in low atmospheric regions. This conclusion confirms the former findings of \citet{Murawski2010}. The patterns of standing waves and magnetic shells are easily seen below the transition region.

Figure~\ref{fig:vert_prof_v} displays the vertical profiles of
$V_\theta$ (left) and $B_\theta$ (right) at $r=0.1~{\rm Mm}$.
The penetration of Alfv\'{e}n waves into the solar corona is seen
in the $V_\theta$ profiles (left). The presented results clearly show that
the wave variables behave differently, namely the gradients in physical 
parameters have stronger effects on $B_\theta$ than on $V_\theta$ because 
the former decays rapidly with height while the latter reaches the solar 
corona; the fact that the Alfv\'{e}n wave variables behave differently is
a well-known phenomenon \citep[e.g.][]{Hollweg1982,Musielak1992,
MusielakMoore1995,Musielak2007,Murawski2010,Routh2010,Perera2015}.  
The presence of gradients is the main physical reason for such a 
different behavior of $V_\theta$ and $B_\theta$, and it also leads to 
wave reflection, which occurs below $y\approx 1.75~{\rm Mm}$ and decays with time; the waves that undergo reflection become trapped 
and this trapping dominates in the upper parts of the solar chromosphere. 
Similar to $V_\theta(r=0.1~\rm{Mm},\textit{y},\textit{t})$, the profiles of $B_\theta$ 
reveal the transient phase, which occurs for $t \lesssim 150~{\rm s}$ but 
later on the oscillations in $B_\theta$ reach a quasi-stationary stage 
(bottom).

The above numerical results have important physical implications,
namely, they show that Alfv\'en waves do lose their identity because 
their wave velocity and magnetic field variables behave differently.
Since the wave velocity variable reaches the corona and the wave 
magnetic field variable cannot, the Alfv\'en wave in the solar corona
does not obey any more its equipartition of energy between the two 
wave variables.  This means that the physical nature of the Alfv\'en 
wave in the solar corona is different than in the solar chromosphere.
The problem is out of the scope of this paper but it does require 
future studies.
\section{Cutoff Periods for Torsional Waves in Thin Flux Tubes}
To compare the results of our numerical simulations to the 
analytically determined conditions for the propagation of torsional Alfv\'en 
waves in a solar magnetic flux tube, we follow \citet{Musielak2007} and \citet{Routh2010}.  With the wave variables $\bm{\vec v}$ = $v_{\theta}(r,y,t) 
\hat \theta$ and $\bm{\vec b}$ = $b_{\theta} (r,y,t) \hat \theta$, we follow \citet{Musielak2007} and write the $\theta$-components of the linearized induction equation 
of motion and induction as
     \begin{equation}
\frac{\partial (v_\theta /r)}{\partial t}-\frac{1}{ \rho_h r^2}
\left [ B_{er} (r,y) \frac{\partial}{\partial r} + B_{\rm{e}y} (r,y) 
\frac{\partial} {\partial y} \right ](r b_\theta) = 0\ , 
     \label{moment1}
     \end{equation}
and 
     \begin{equation}
\frac{\partial (r b_{\theta})}{\partial t} - r^2 \left [ B_{er} (r,y)
\frac{\partial}{\partial r} + B_{\rm{e}y} (r,y) \frac{\partial}{\partial y} 
\right ]  (v_\theta / r) = 0\ .
     \label{induc1}
     \end{equation}

According to \citet{Musielak2007}, the thin flux tube approximation 
gives the following relationship:
     \begin{equation}
B_{\rm{e}r}(r,y) = - \frac{r}{2} {{d B_{\rm{e}y}} \over {dy}}  \ ,
     \label{thin1}
     \end{equation}
which allows writing Equations (11) and (12) as 
     \begin{equation}
\frac{\partial{v_{\theta}}}{\partial t} + \frac{1}{2 \rho_h} 
\left ( r \frac{\partial {b_{\theta}}}{\partial r} + b_{\theta} 
\right ) {{d B_{\rm{e}y}} \over {dy}} - \frac{B_{\rm{e}y}} {\rho_h} 
\frac{\partial b_{\theta}} {\partial y} = 0\ ,
     \label{moment2}
     \end{equation}
and
     \begin{equation}
\frac{\partial {b_{\theta}}}{\partial t} + \frac{1}{2} \left ( 
r \frac{\partial {v_{\theta}}}{\partial r} - v_{\theta} \right ) 
{{d B_{\rm{e}y}} \over {dy}} - B_{\rm{e}y} \frac{\partial v_{\theta}} 
{\partial y} = 0\ .
     \label{induc2}
     \end{equation} 

Assuming the tube being in temperature equilibrium with its 
surroundings, \citet{Routh2010} used horizontal pressure
balance and obtained the following wave equations for torsional
Alfv\'en waves:
         \begin{equation} 
\frac{\partial^{2} v_{\theta}}{\partial t^2} - c_{A}^{2} 
\frac{\partial^{2} v_{\theta}} {\partial y^2} + 
\frac{c_{A}^{2}} {2H } \frac{\partial v_{\theta}} {\partial y} 
- \frac{c_{A}^{2} } {4H^2 }\left(\frac{1}{4} + {{dH} \over 
{dy}} \right) v_{\theta} = 0\ , 
         \label{weq1}
         \end{equation}
and
        \begin{equation}
\frac{\partial^{2}b_\theta} {\partial t^2} - c_{A}^{2}  
\frac{\partial^{2} b_{\theta}}{\partial y^2} - {{c_{A}^{2}}
\over {2 H}} \left ( 1 + \frac {4H}{c_{\rm{A}}} {{d c_{\rm{A}}} \over 
{dy}} \right ) \frac{\partial b_\theta}{\partial y}
-\frac{c_{A}^{2}}{4H^2} \left ( \frac{1}{4} + \frac{2H} 
{c_{\rm{A}}} {{d c_{\rm{A}}} \over {dy}} - {{d H} \over {dy}} \right ) 
b_{\theta} = 0\\,
        \label{weq2}
        \end{equation}
\noindent
where $H(y) =  c_{\rm{s}}^2(y) / \gamma g$ is the pressure scale 
height with $c_{\rm{s}}(y)$ being the sound speed, and 
$c_{\rm{A}}(y) = B_{\rm{e}y}(y) / \sqrt{ \rho_h(y)}$ is the Alfv\'en velocity. 

As demonstrated by \citet{Routh2010}, the critical frequencies are given by
\begin{equation}
\Omega_{\rm{cr},v}^{2}(y) = \frac{1}{2}\left[\frac{1}{2}\left({\frac{dc_{\rm{A}}}
{dy}}\right)^2-c_{\rm{A}}\left(\frac{d^2c_{\rm{A}}}{dy^2}\right)\right],
\end{equation}
and
\begin{equation}
\Omega_{\rm{cr},b}^{2}(y) = \frac{1}{2}\left[\frac{1}{2}\left({\frac{dc_{\rm{A}}}
{dy}}\right)^2+c_{\rm{A}}\left(\frac{d^2c_{\rm{A}}}{dy^2}\right)\right],
\end{equation}
and the resulting turning-point frequencies become
\begin{equation}
\Omega_{\rm{tp},v}^{2}(y) = \Omega_{\rm{cr},v}^{2}(y) + \frac{1}{4t^2_{ac}(y)},
\end{equation}
and
\begin{equation}
\Omega_{\rm{t}p,b}^{2}(y) = \Omega_{\rm{cr},b}^{2}(y) + \frac{1}{4t^2_{ac}(y)},
\end{equation}
where $t_{ac}$ is actual wave travel time expressed by
\begin{equation}
t_{\rm{ac}}=\int^{y}_{y_b}\frac{d\widetilde{y}}{c_{\rm{A}}(\widetilde{y})},
\end{equation}
with $y_b$ being an atmospheric height at which the wave is initially 
generated.
Finally, the cutoff frequency for torsional Alfv\'en waves propagating 
in a thin and non-isothermal magnetic flux tube embedded in the solar
atmosphere is 
     \begin{equation}
\Omega_{\rm{cut},\tau} (y) = {\max} [\Omega_{\rm{tp},\tau,v}(y), 
\Omega_{\rm{tp},\tau,b} (y)]\ .
     \label{cutoff}
     \end{equation}
Having obtained the turning-point frequencies and the cutoff frequency,
we may define the turning-point periods as 
\begin{equation}
P_{\rm{tp},v}=\frac{2 \pi}{\Omega_{\rm{tp},v}}\,, \hspace{3mm}
P_{\rm{tp},b}=\frac{2 \pi}{\Omega_{\rm{tp},b}},
\end{equation}
and the cutoff period is 
\begin{equation}
P_{\rm{cut}}=\frac{2 \pi}{\Omega_{\rm{cut}}}.
\end{equation}
\begin{figure}[h!]
\centerline{\includegraphics[width=0.5\textwidth,clip=]{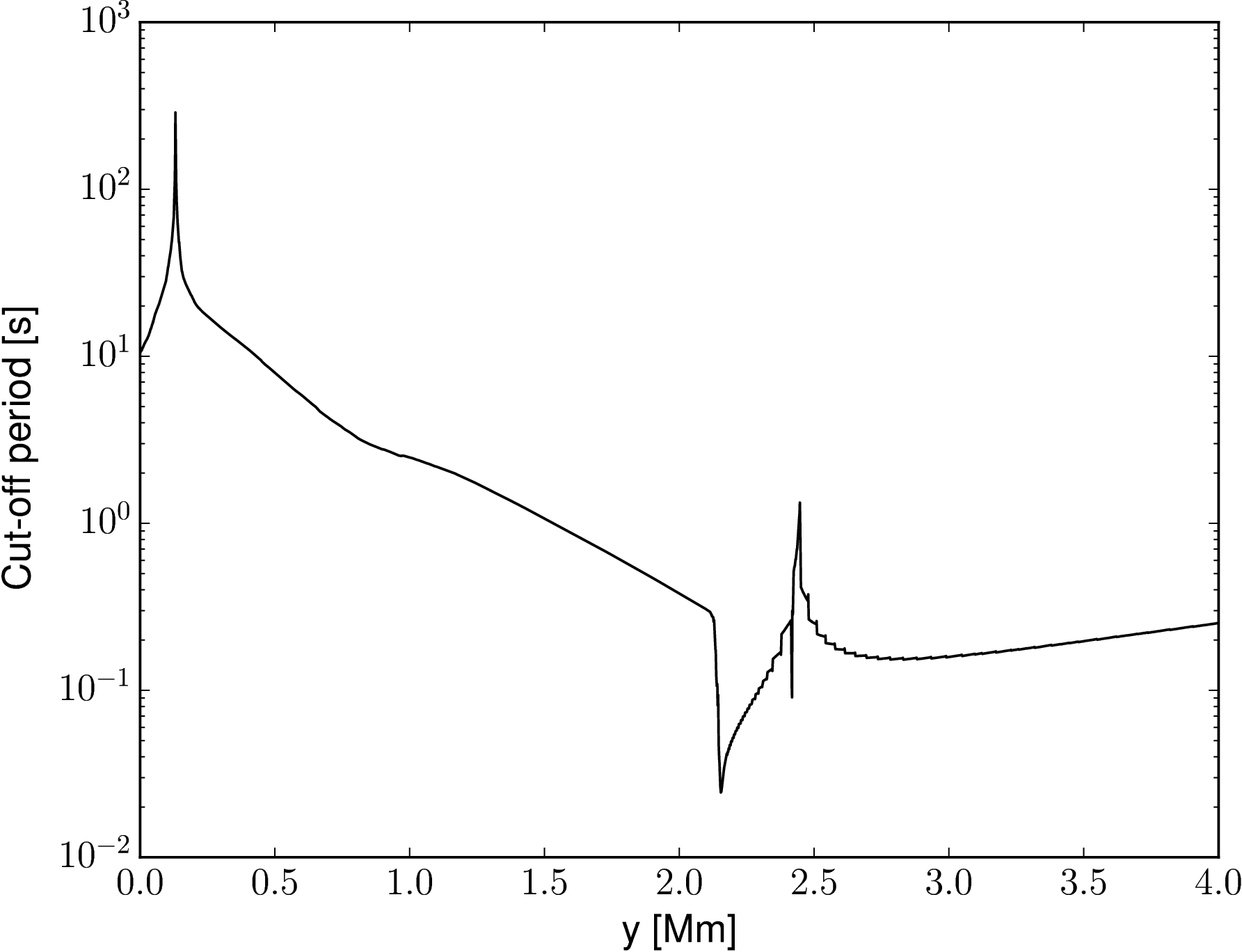}}
\caption{
Cutoff period for Alfv\'{e}n waves \textit{vs} height $y$ calculated from Equations~(23) and (25).}
\label{fig:cutoff_theory}
   \end{figure}
\section{Discussion of Analytical and Numerical Results}
Figure~\ref{fig:cutoff_theory} displays the cutoff period, $P_{cut}$, \textit{vs} height $y$. 
At $y=0$, which is the bottom of the photosphere, $P_{\rm{cut}}=10$ s. Higher up it grows, 
reaching its maximum value of about $300$ s at $y\approx 0.2$ Mm, and subsequently it 
falls off with height to its global minimum of about $0.03$ s, which occurs at the 
solar transition region.  In the solar corona $P_{\rm{cut}}$ increases again with $y$ but 
its values are lower than $1$ s.  This general trend of $P_{cut}(y)$ is consistent 
with that found by \citet[]{Murawski2010}, however, the exact values of $P_{\rm{cut}}(y)$ 
are different in the two cases.  In particular, \citet[]{Murawski2010} found that for 
the straight vertical equilibrium magnetic field $P_{\rm{cut}}$ was higher than $40$ s, 
while Figure~\ref{fig:cutoff_theory} reveals the corresponding values being a fraction 
of $1$ s. This clearly shows how strongly the cutoff depends on the geometry and 
strength of the background magnetic field. 

Our numerical results show that Alfv\'en waves - or at least their wave velocity 
perturbation - penetrate the transition region, whereas the magnetic field perturbation
does not, as shown in Figures~\ref{fig:vert_prof_v} and \ref{fig:spat_prof_v}. 
The analytical results imply the same as they also show that the two wave variables
behave differently. However, there are some discrepancies between the analytical 
and numerical results because the thin flux tube approximation, which is the basis 
for obtaining the cutoff frequency, does not apply to our model of the solar corona
because the tube becomes thick and the condition given by Equation (13) is not satisfied 
any longer.  The agreement between the analytical theory and our numerical model is
better in the solar chromosphere where the tube remains thin.  

Because of the different behavior of the two wave variables, one may look directly at
the values of the turning-point frequencies given by Equations (20) and (21) for each wave
variable.  Since in the photosphere for $0.2~{\rm Mm}<y<2.5~{\rm Mm}$, $P_{\rm{tp},b}=0$, a 
signal in magnetic field is not able to penetrate this region, which explains why 
$B_{\theta}$ is absent in the solar corona (Figure~\ref{fig:vert_prof_v}, right); 
this is an important result, which clearly shows the connections between the 
turning-point frequencies and the different behavior of the two wave variables. However, 
a signal can be seen in this region in $V_{\theta}$ (Figure~\ref{fig:vert_prof_v}, left). 
Indeed, for $P_{\rm d} =35~\rm s$ the condition of $P<P_{tp,v}$ is satisfied only up 
to $y\approx0.3~\rm Mm$, while for $y\approx0.1~\rm Mm$ $P<P_{\rm{tp},b}$. For larger 
values of y, we have $P>P_{\rm{tp},bv}$ and the Alfv\'{e}n waves become evanescent.

\begin{figure} 
                                % includes the two top panels 
   \centerline{\hspace*{0.015\textwidth}
               \includegraphics[width=0.5\textwidth,clip=]{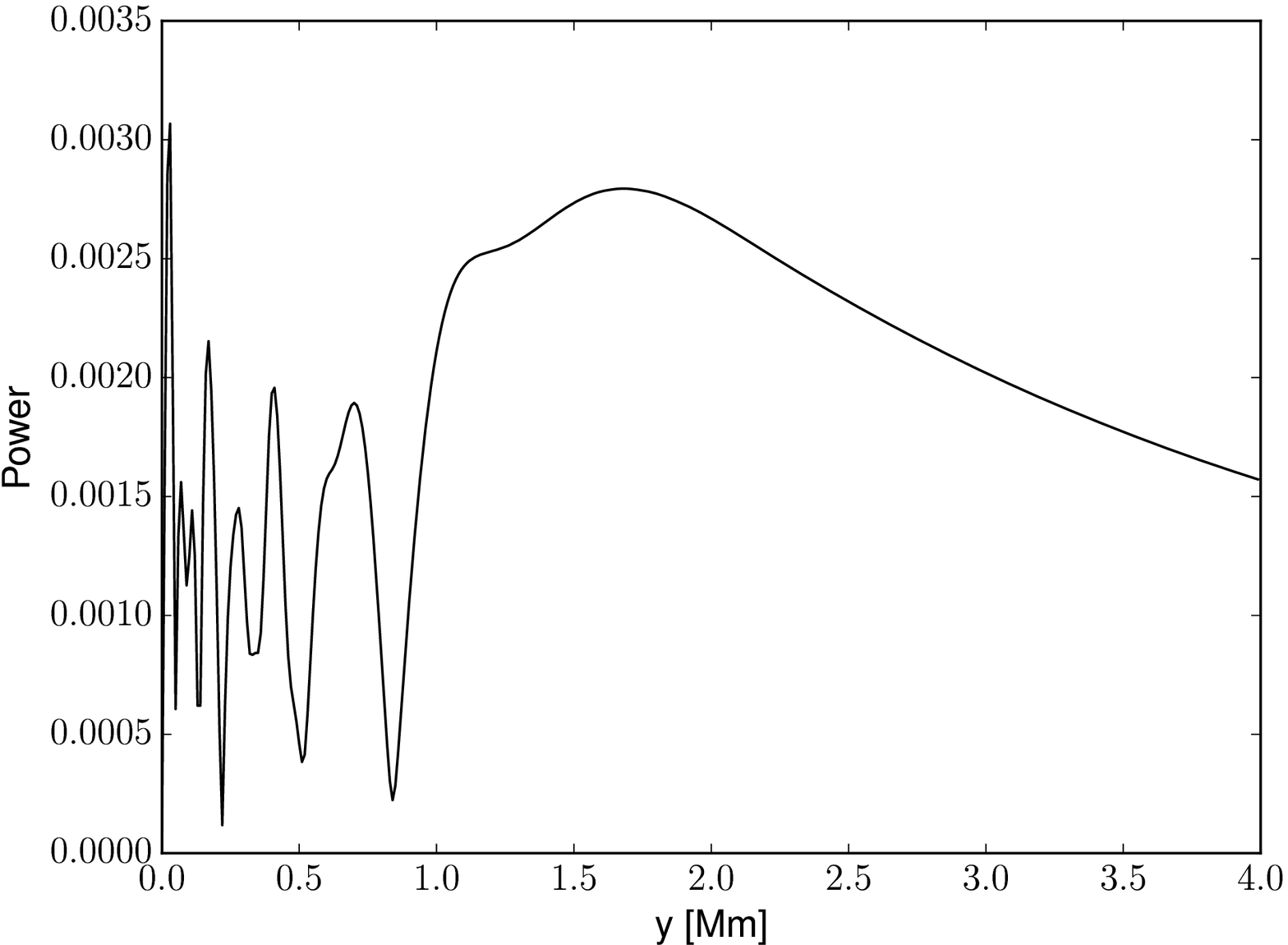}
               \hspace*{0.0\textwidth}
               \includegraphics[width=0.5\textwidth,clip=]{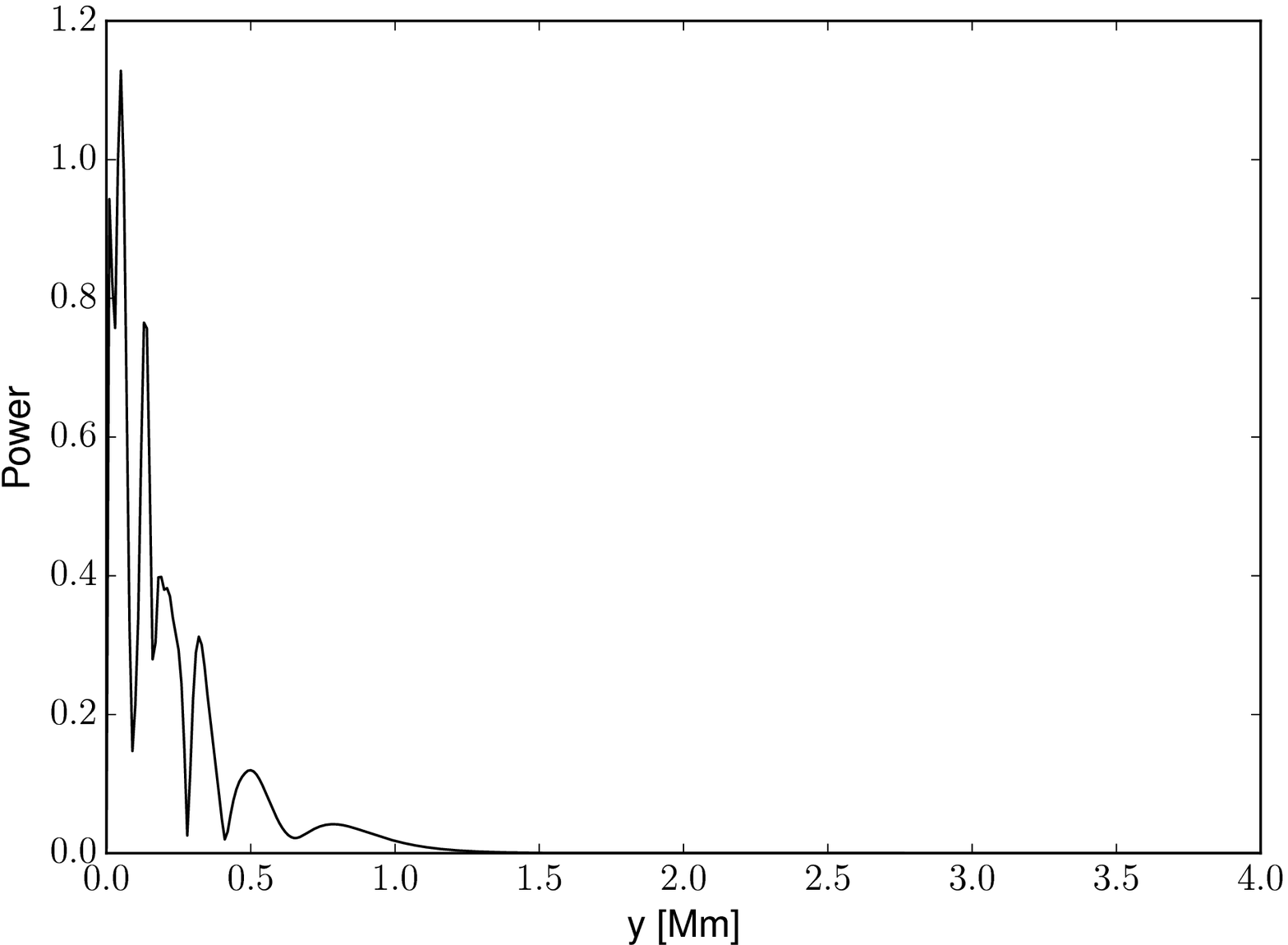}
              }        
\caption{
Spectral power (in arbitrary units) of Alfv\'{e}n waves for wave period $P=35\rm~s$ obtained 
from the time signatures of Figure~\ref{fig:vert_prof_v} for $r=0.1~{\rm Mm}$ and $t=250~\rm s$; 
for $V_\theta$ (left) and $B_\theta$ (right).}
\label{fig:power_of_wave}
   \end{figure}

Figure~\ref{fig:power_of_wave} illustrates spectral power corresponding to $P_{\rm d}
=35~\rm s$ for $V_{\theta}$ (left) and $B_{\theta}$ (right). A sudden fall-off of power 
at $y\approx0.2~\rm Mm$ corresponds to energy being spent on excitation of a non-zero 
signal in $B_{\theta}$. Thus energy goes from $V_{\theta}$, in which it was originally 
generated, into $B_{\theta}$.  We point out that for linear Alfv\'{e}n waves 
propagating in homogeneous media, there is a perfect equipartition of the wave kinetic 
energy associated with $V_{\theta}$ and the wave magnetic energy associated with 
$B_{\theta}$; our results show that this perfect equipartition of wave energy is
strongly violated in inhomogeneous media (see our comments at the end of Section 3). 

In the solar corona the value of the $P_d$ is larger then a local value of $P_{\rm{tp},v}$ and $P_{\rm{tp},b}$ so the
waves are evanescent in this region.  However, our numerical results show a tunneling 
of $V_{\theta}$ (left) to the solar corona with the signal in $B_{\theta}$ (right) falling 
off with height more strongly, with and $B_{\theta}$ being practically zero above $y=1.5~\rm Mm$.

  \begin{figure}[H]   
                                % includes the two top panels 
   \centerline{\hspace*{0.015\textwidth}
               \includegraphics[width=0.515\textwidth,clip=]{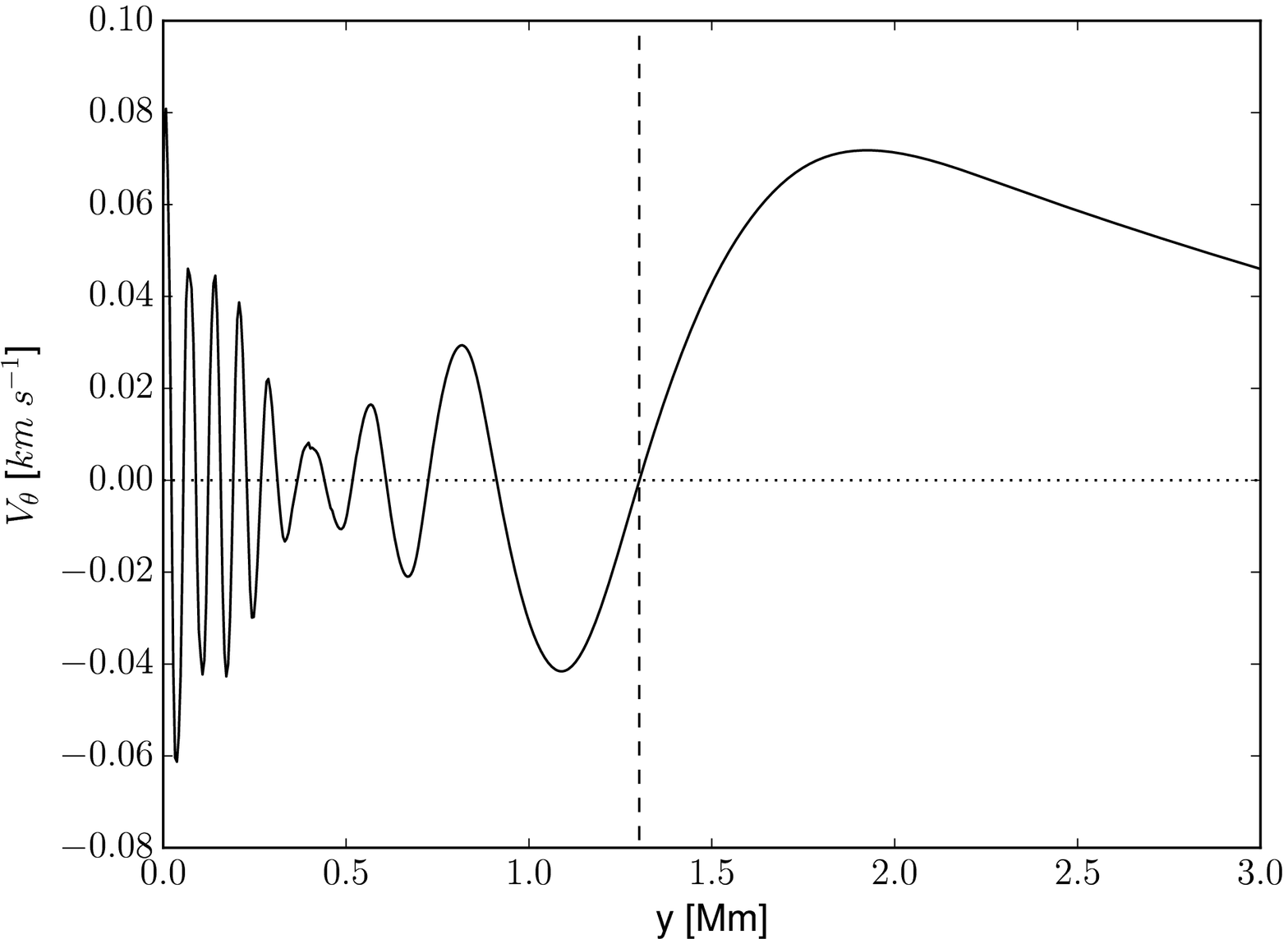} 
               \hspace*{0.0\textwidth}
               \includegraphics[width=0.515\textwidth,clip=]{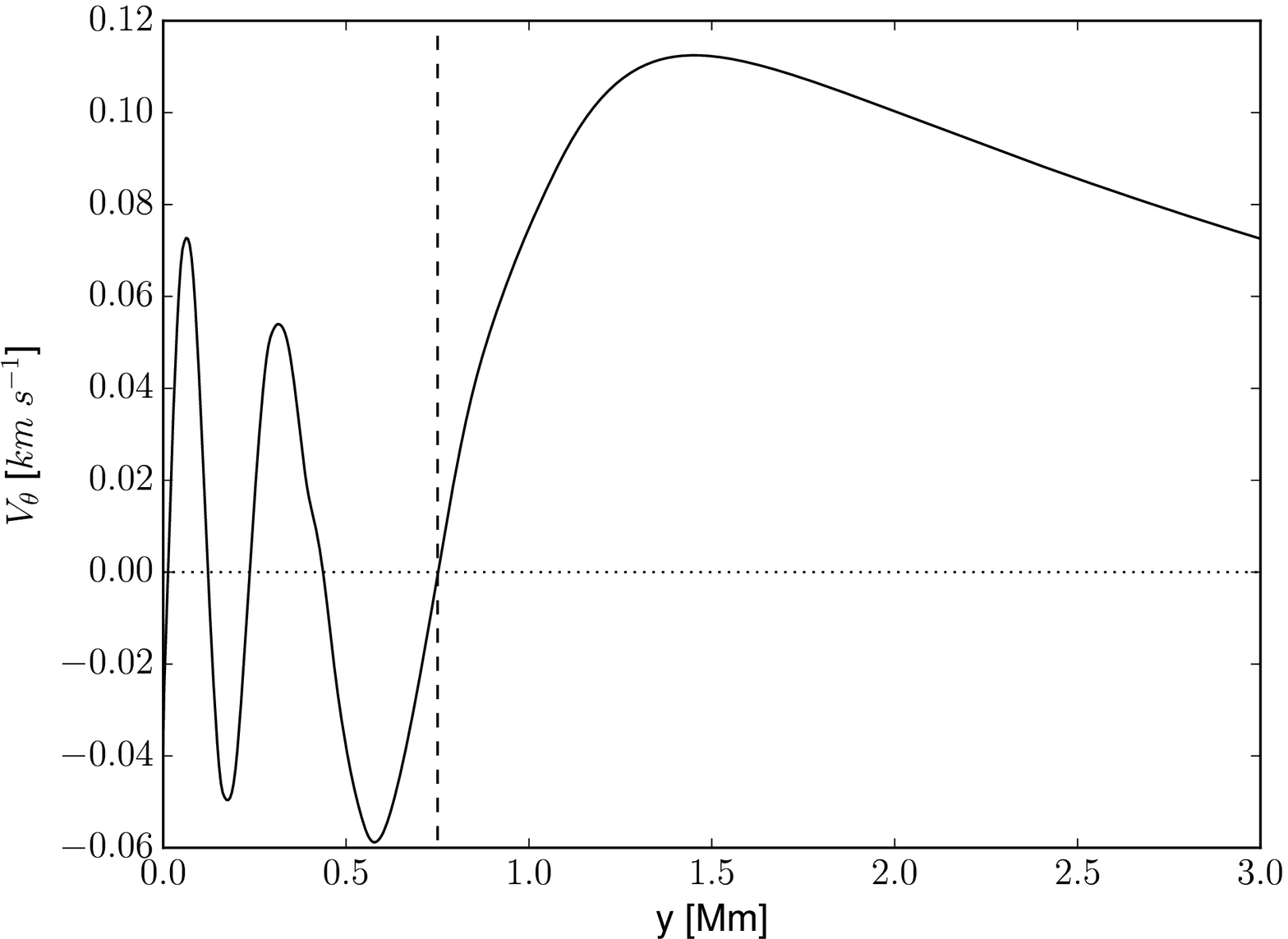}
              }
     \vspace{-0.35\textwidth}   % Shift close to the panel top 
     \centerline{\Large \bf     % Includes the labels (here needs the color 
                                %   package, see beginning of this file)
      \hspace{0.0 \textwidth}  \color{white}{(a)}
      \hspace{0.415\textwidth}  \color{white}{(b)}
         \hfill}
     \vspace{0.31\textwidth}    % Shift back to the panel bottom 
   \centerline{\hspace*{0.015\textwidth}
               \includegraphics[width=0.515\textwidth,clip=]{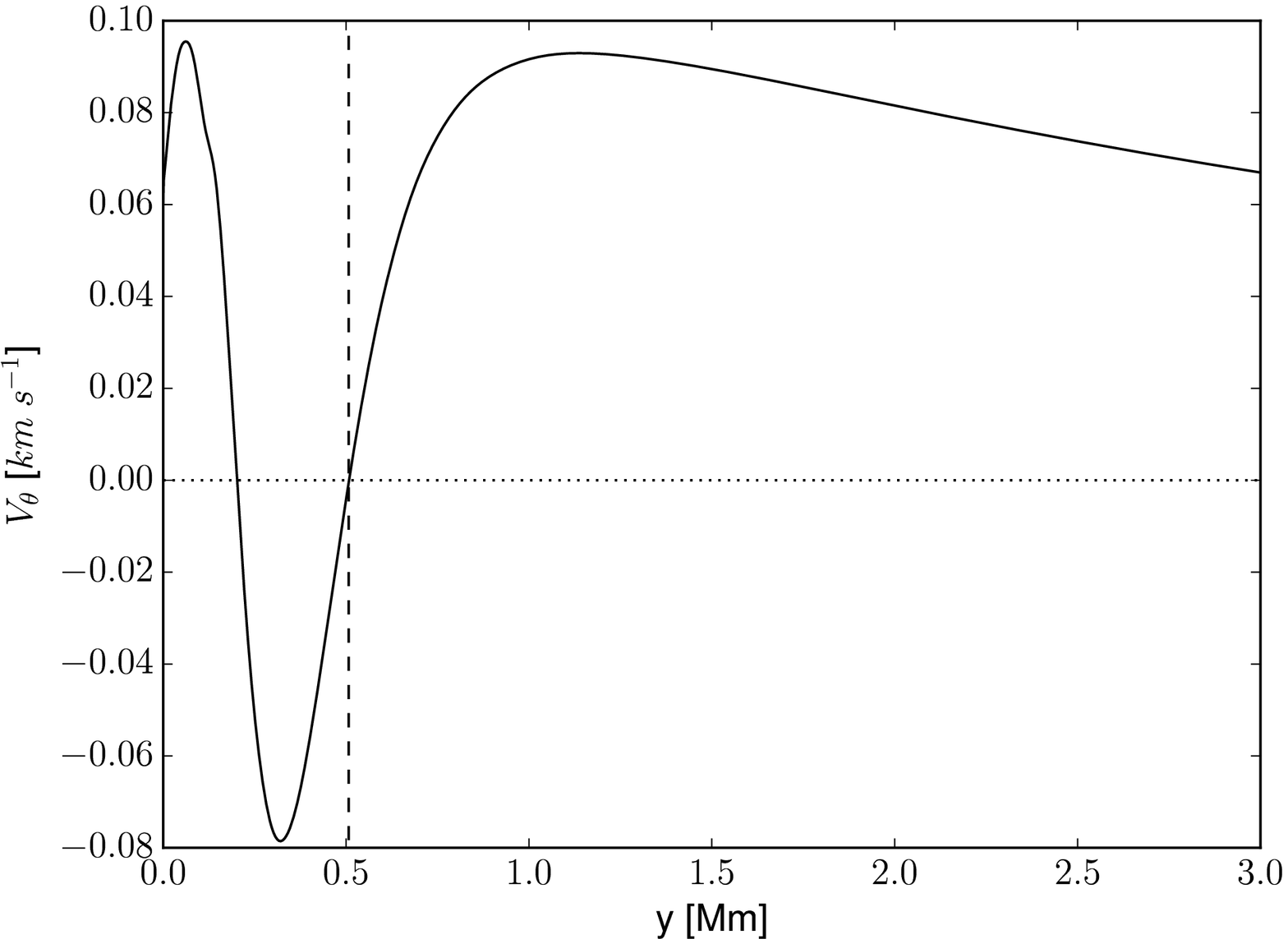} 
               \hspace*{0.0\textwidth}
               \includegraphics[width=0.515\textwidth,clip=]{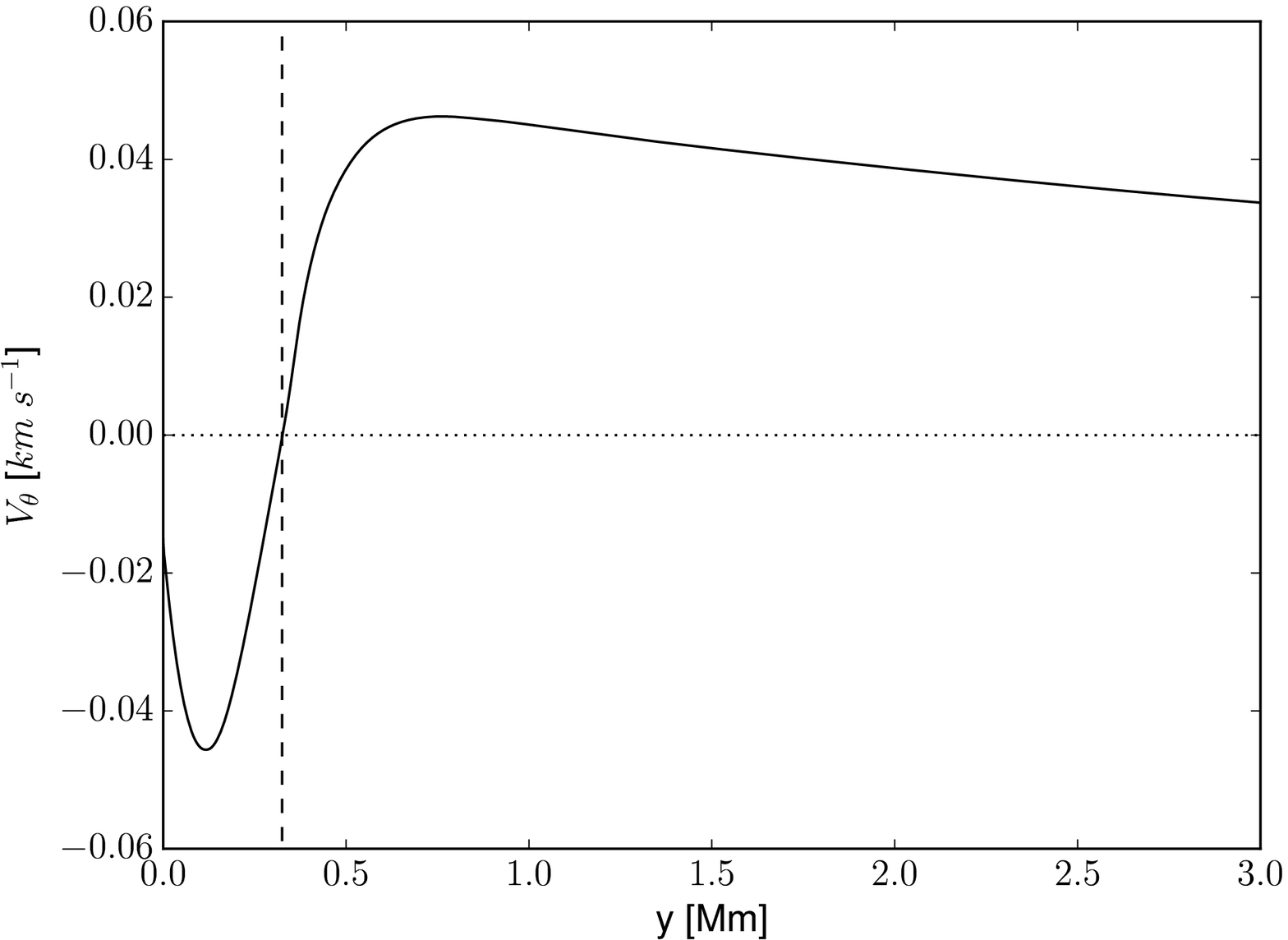} 
              }
     \vspace{-0.35\textwidth}   % Shift close to the panel top 
     \centerline{\Large \bf     % Includes the labels (here needs the color package)
      \hspace{0.0 \textwidth} \color{white}{(c)}
      \hspace{0.415\textwidth}  \color{white}{(d)}
         \hfill}
     \vspace{0.31\textwidth}    % Shift back to the panel bottom 
              
 \caption{Plots of $V_\theta(r=0.1~\rm Mm,\rm y)$ \textit{vs} the atmospheric height $\rm y$ with 
	the locations of the turning point depicted by vertical dashed lines for: $P_{\rm d}$=35~s (top left), $P_{\rm d}$=50~s (top right), $P_{\rm d}$=100~s (bottom left) and $P_{\rm d}$=200~s (bottom right).}
  \label{fig:waves_vx}
   \end{figure}
   
     \begin{figure}
                                % includes the two top panels 
   \centerline{\hspace*{0.015\textwidth}
               \includegraphics[width=0.515\textwidth,clip=]{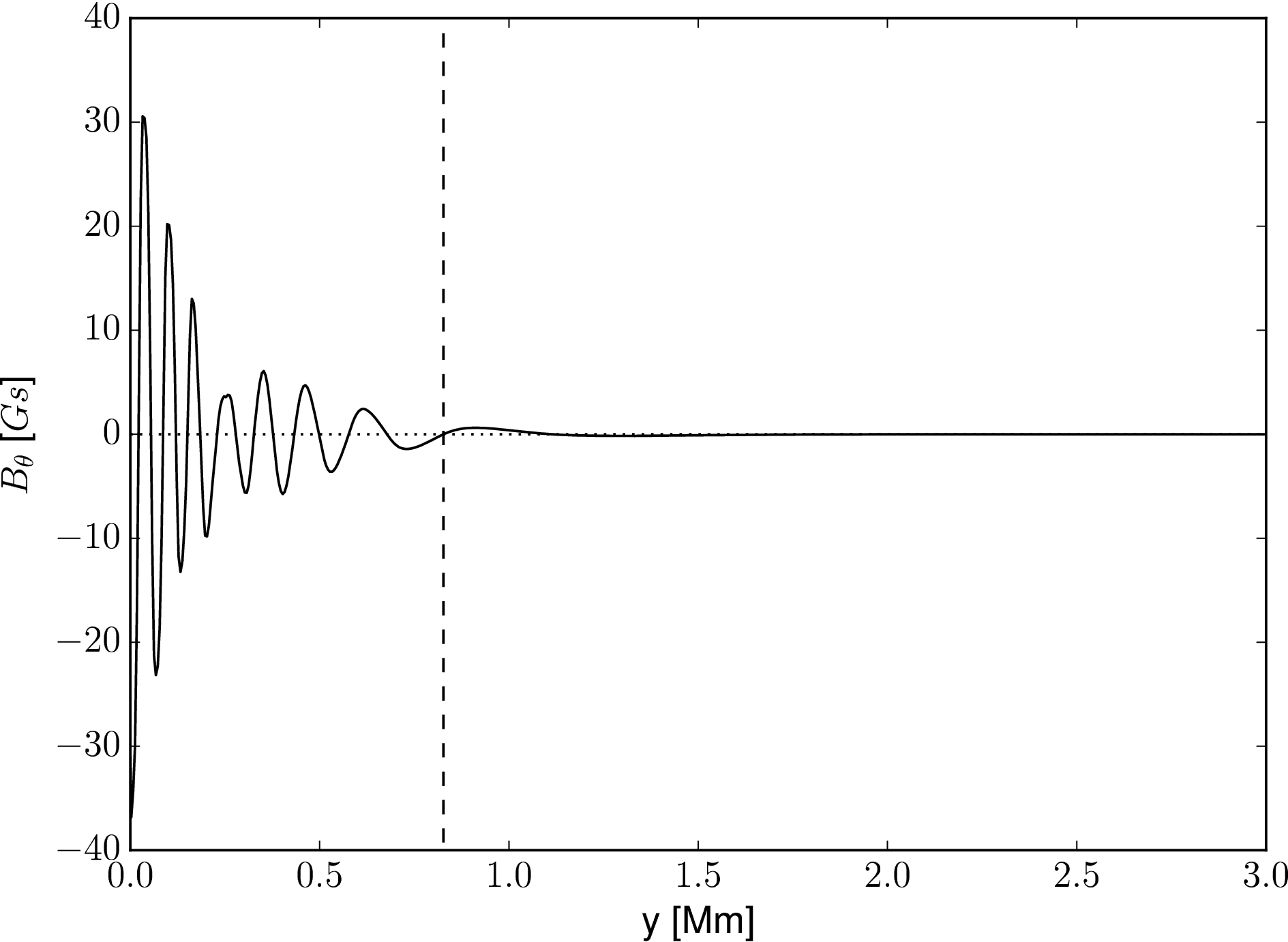}  
               \hspace*{0.0\textwidth}
               \includegraphics[width=0.515\textwidth,clip=]{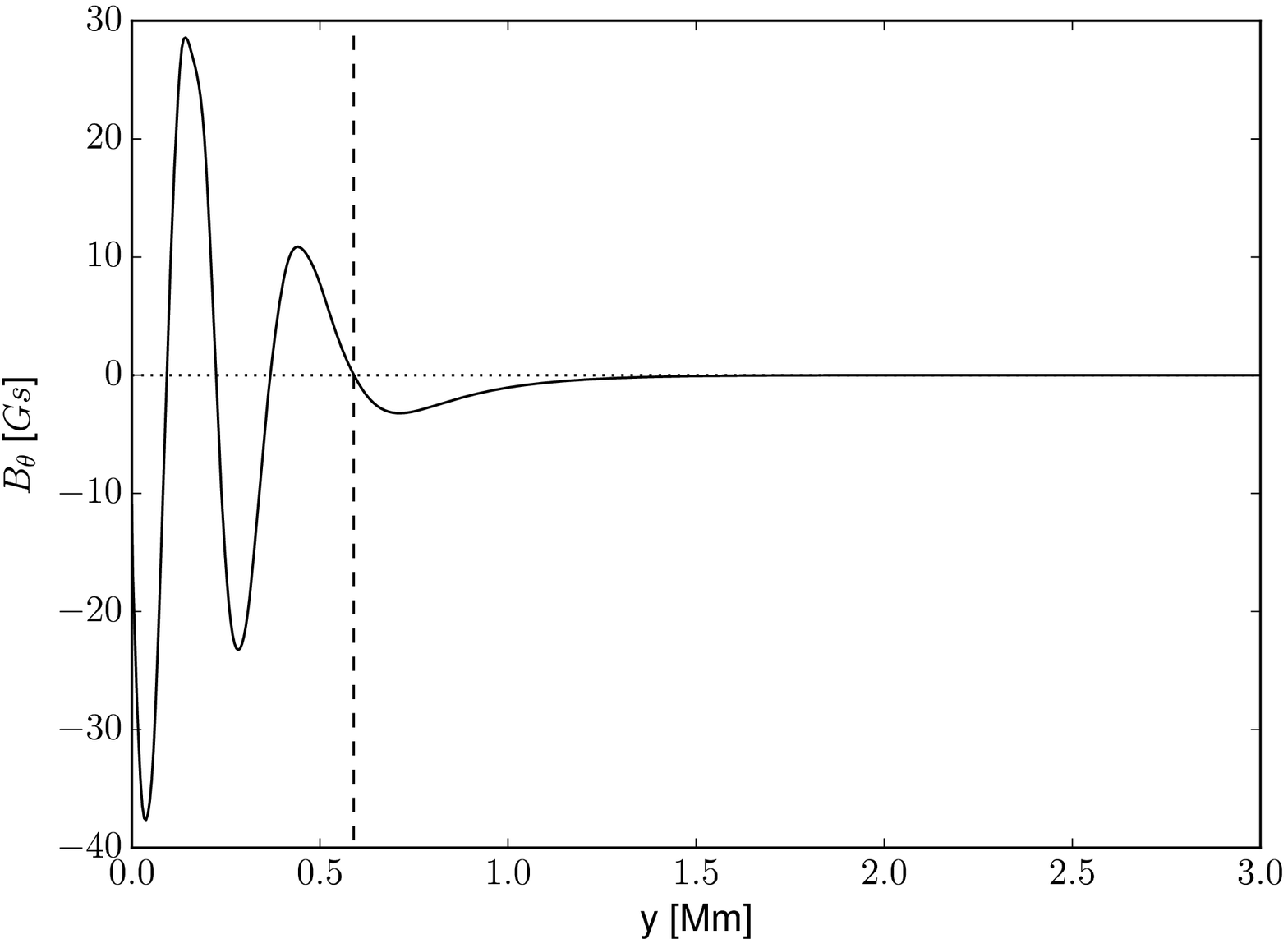} 
              }
     \vspace{-0.35\textwidth}   % Shift close to the panel top 
     \centerline{\Large \bf     % Includes the labels (here needs the color 
                                %   package, see beginning of this file)
      \hspace{0.0 \textwidth}  \color{white}{(a)}
      \hspace{0.415\textwidth}  \color{white}{(b)}
         \hfill}
     \vspace{0.31\textwidth}    % Shift back to the panel bottom 
   \centerline{\hspace*{0.015\textwidth}
               \includegraphics[width=0.515\textwidth,clip=]{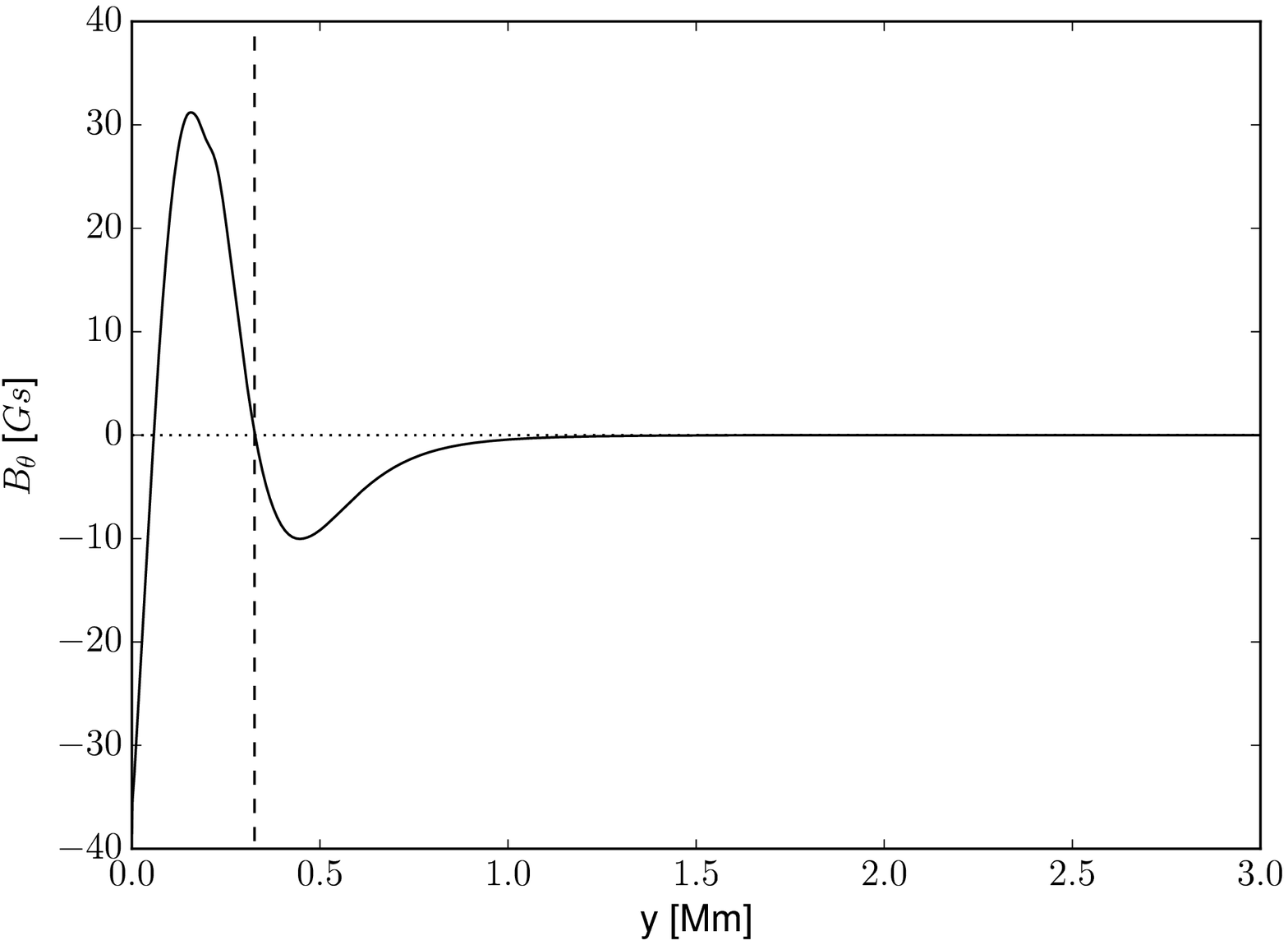}  
               \hspace*{0.0\textwidth}
               \includegraphics[width=0.515\textwidth,clip=]{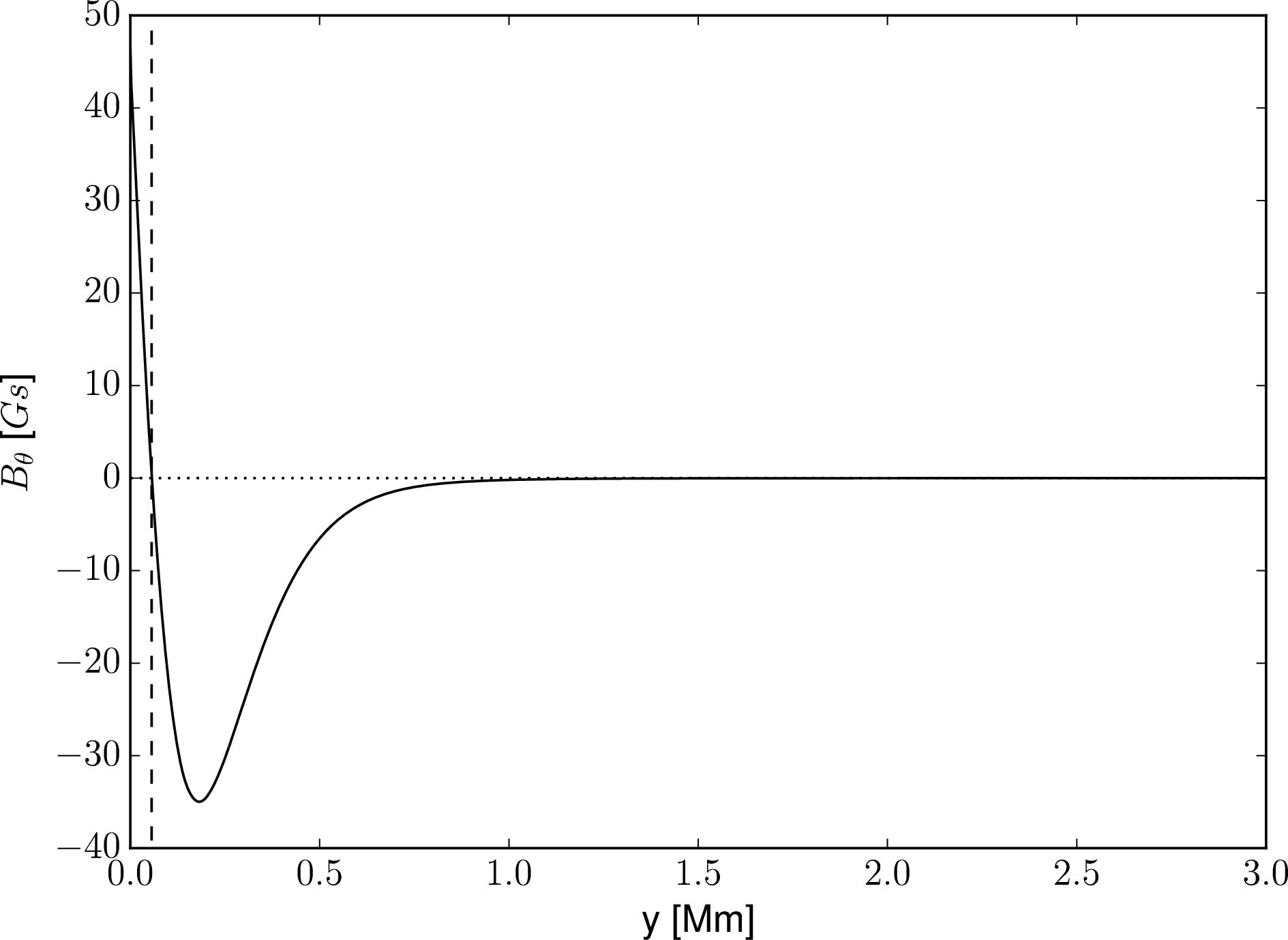} 
              }
     \vspace{-0.35\textwidth}   % Shift close to the panel top 
     \centerline{\Large \bf     % Includes the labels (here needs the color package)
      \hspace{0.0 \textwidth} \color{white}{(c)}
      \hspace{0.415\textwidth}  \color{white}{(d)}
         \hfill}
     \vspace{0.31\textwidth}    % Shift back to the panel bottom 
              
\caption{Plots of $B_\theta(r=0.1~\rm Mm,\rm y)$ \textit{vs} the atmospheric height $\rm y$ with the 
locations of the turning point depicted by vertical dashed lines for: $P_{\rm d}$=35~s (top left), $P_{\rm d}$=50~s (top right), $P_{\rm d}$=100~s (bottom left) and $P_{\rm d}$=200~s (bottom right).}
\label{fig:waves_bx} 
   \end{figure}

Figure~\ref{fig:waves_vx} shows the vertical profiles of $V_\theta(r=0.1~\rm Mm,\rm y)$ for several values
of $P_{\rm d}$. Location of the last zero of $V_\theta$ is assumed as the $y$-position of a turning point. 
Similar profiles are drawn for $B_\theta(r=0.1~\rm Mm,y)$, (see Figure~\ref{fig:waves_bx}). Note that 
oscillations take place until the zero of $V_\theta$ and higher up $V_\theta$ becomes evanescent as 
it falls off with altitude above its local maximum. The oscillatory behavior of $B_\theta$ until 
the last zero and its subsequent strong fall-off are clearly seen.

\begin{figure}[ht] 
   \centerline{\includegraphics[width=0.5\textwidth,clip=]{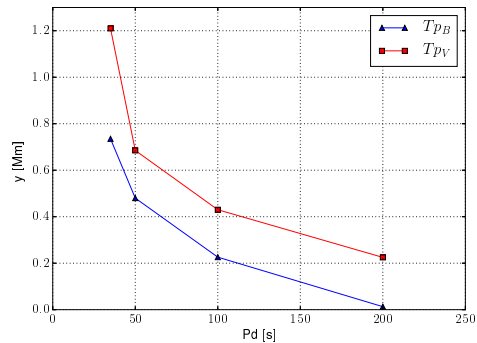}
              }
\caption{
The vertical location of the turning point for $B_\theta$ (triangles) and $V_\theta$ 
(squares) \textit{vs} driving period $P_{\rm d}$.}
\label{fig:pd_vs_t_tp}
   \end{figure}

The turning points for $V_\theta$ and $B_\theta$, which separate oscillatory behavior of the wave
variables from non-oscillatory behavior, are illustrated in Figure~\ref{fig:pd_vs_t_tp}.  Note that 
the turning points are located at lower heights for larger values of $P_d$ with the falling-off 
trend confirming the numerical findings of \citet{Perera2015}, and that the turning points 
for $B_\theta$ are located at lower altitudes (see Figure~\ref{fig:pd_vs_t_tp}). There is strong consistency between our conclusions based on the cutoff and on the concept of turning points, and the reason is that all these two concepts 
are related to each other \citep[\textit{e.g.}][]{Musielak2006}.
\section{Summary and Concluding Remarks}
In this paper, we simulate numerically Alfv\'{e}n waves which are driven by a periodic 
driver operating in the solar photosphere 100 km below the solar surface.  Our findings 
can be summarized as follows
\begin{description}
\item[\rm{(a)}]A periodic driver located in the solar photosphere excites torsional Alfv\'{e}n 
waves that propagate along an axi-symmetric magnetic flux tube embedded in the solar 
chromosphere and corona.
\item[\rm{(b)}]It is shown that the azimuthal component of velocity $V_{\theta}$ and the 
azimuthal component of $B_{\theta}$ behave very differently in the solar atmosphere;
$V_{\theta}$ propagates in the solar corona, while $B_{\theta}$ exhibits a strong fall-off 
trend with height.
\item[\rm{(c)}]This different behavior of the two wave variables can be explained by the fact
that the variables for Alfv\'{e}n waves have different turning point frequencies, and that 
these frequencies are used to define the cutoff frequency (or period), which sets up the 
conditions for the wave propagation in the solar atmosphere.
\item[\rm{(d)}]The presence of the cutoff for Alfv\'{e}n waves implies that the waves with 
periods close to the cutoff are being reflected and that interaction between incoming 
and reflected waves leads to the formation of standing wave patterns.
\item[\rm{(e)}] Our numerical and analytical results demonstrate that the Alfv\'en wave loses
its identity while traveling in the solar atmosphere because the wave velocity 
perturbation reaches the solar corona but the wave magnetic field perturbation does not;
as a result, the wave variables do not satisfy equipartition of energy and 
thus the Alfv\'en wave has a different nature in the upper atmospheric layers.
\end{description}

The obtained results extend the previous work of \citet{Murawski2010} and 
\citet{Perera2015} to describe torsional Alfv\'{e}n waves propagating along solar magnetic
flux tubes.  From a mathematical point of view, the previous and current work is
different, namely, our study is mathematically more complex.
However, from a physical point of view, there are some  differences and
similarities between Alfv\'{e}n waves studied in the previous work and the torsional 
Alfv\'{e}n waves investigated here. 
In particular, the previous work by \citet{Musielak2007} and \citet{Routh2010}
was used to obtain the cutoff frequency for torsional Alfv\'en waves propagating 
along a thin and non-isothermal solar magnetic flux tube. The approach formally 
takes into account both $B_{\rm{e}r} (r,y)$ and $B_{\rm{e}y} (r,y)$, however, only within the
thin flux tube approximation.  Our analytical results in Section 4 show that
the turning-point frequencies are the same as those obtained by \citet{Murawski2010},
who considered regular Alfv\'en waves propagating along a uniform background 
magnetic field, however, the cutoff wave periods are different.
The results presented here clearly show that the wave variables behave differently 
and that the atmospheric inhomogeneities have stronger effects on the wave magnetic 
variable than on the wave velocity variable; again, this result is consistent with
that obtained previously by \citet{Murawski2010} and \citet{Perera2015}.  Moreover, 
we demonstrate that one of the best ways to account for these differences 
is to use the turning point frequencies because their values uniquely determine 
the behavior of each wave variable in the solar atmosphere, and because they are 
used to define the cutoff frequency (or period) for Alfv\'en waves.  It is also 
important to emphasize that the different behavior of the Alfv\'{e}n wave variables
in inhomogeneous media leads to violation of the equipartition of the wave kinetic and 
magnetic energies in Alfv\'{e}n waves and that this fact makes a clear distinction
between Alfv\'{e}n waves propagating in homogeneous and inhomogeneous media.
   
%%%%%%%%%%%%%%%%%%%%%%%%%%%%%%%%%%%%%%%%%%%%%%%%%%%%%%%%%%%%%%%%%%%%%%%%%%%
%% Appendix
%
% \appendix   

%%%%%%%%%%%%%%%%%%%%%%%%%%%%%%%%%%%%%%%%%%%%%%%%%%%%%%%%%%%%%%%%%%%%%%%%%%%
%% Acknowledgements
%
 \begin{acks} 
The authors would like to express their thanks to the referee for his/her 
comments and suggestions that allowed us to significantly improve the 
original manuscript.  This work was supported by the project from Polish 
National Science Foundation under the grant No. 2014/15/B/ST9/00106 (D.W., 
K.M., Z.E.M. and P.K.), by NSF under the grant AGS 1246074 (Z.E.M. and K.M.),
and by the Alexander von Humboldt Foundation (Z.E.M.). Numerical studies 
have been performed on the LUNAR cluster at Institute of Mathematics of 
University of M. Curie-Sklodowska in Lublin, Poland.
 \end{acks}
 
\begin{disc}
The authors declare that they have no conflicts of interest.
\end{disc}

%%% %%%%%%%%%%%%%%%%%%%%%%%%%%%%%%%%%%%%%%%%%%%%%%%%%%%%%%%%%%%
%% Bibliography
%
% Using BibTeX
%

\end{article} 

\begin{thebibliography}{99}

\bibitem[\protect\citeauthoryear{An et al.}{1989}]{An1989}
An C.H., Musielak Z.E., Moore R.L., Suess S.T.: 1989, \textit{Astrophys. J.}, {\bf 345}, 597. \href{http://adsabs.harvard.edu/abs/1989ApJ...345..597A}{ADS}. \href{http://dx.doi.org/10.1086/167933}{DOI}

\bibitem[\protect\citeauthoryear{Avrett and Loeser}{2008}]{Avrett2008}
Avrett E.H., Loeser R.: 2008, \textit{Astrophys. J. Suppl. S}, {\bf 175}, 1. \href{http://adsabs.harvard.edu/abs/2008ApJS..175..229A}{ADS}. \href{http://dx.doi.org/10.1086/523671}{DOI}.

\bibitem[\protect\citeauthoryear{Banerjee, Hasan and Christensen-Dalsgaard}{1998}]{Banerjee1998}
Banerjee D., Hasan, S.S., Christensen-Dalsgaard, J.: 1998, \textit{New Eyes to See Inside the Sun and Stars}, {\bf 185}, 423. \href{http://adsabs.harvard.edu/abs/1998IAUS..185..423B}{ADS}.

\bibitem[\protect\citeauthoryear{Bemporad and Abbo}{2012}]{Bemporad2012}
Bemporad A., Abbo L.: 2012, \textit{Astrophys. J.}, {\bf 751}, 2. \href{http://adsabs.harvard.edu/abs/2012ApJ...751..110B}{ADS}. \href{http://dx.doi.org/10.1088/0004-637X/751/2/110}{DOI}.

\bibitem[\protect\citeauthoryear{Chmielewski et al.}{2013}]{Chmielewski2013}
Chmielewski P., Srivastava A.K., Murawski K., Musielak Z.E.: 2013, \textit{Mon. Not. R. Astron. Soc.}, {\bf 428}, 40. 
\href{http://adsabs.harvard.edu/abs/2013MNRAS.428...40C}{ADS}. \href{http://dx.doi.org/10.1093/mnras/sts009}{DOI}.

\bibitem[\protect\citeauthoryear{Chmielewski et al.}{2014}]{Chmielewski2014}
Chmielewski P., Srivastava A.K., Murawski K., Musielak Z.E.: 2014, \textit{Acta Phys. Pol. A}, {\bf 125}, 1. 
\href{http://adsabs.harvard.edu/abs/2014AcPPA.125..158C}{ADS}. 

\bibitem[\protect\citeauthoryear{Cirtain}{2007}]{Cirtain2007} 
Cirtain J.W., et al.: 2007, \textit{Science}, {\bf 318}, 1580.
\href{http://adsabs.harvard.edu/abs/2007Sci...318.1580C}{ADS}. \href{http://dx.doi.org/10.1126/science.1147050}{DOI}.

\bibitem[\protect\citeauthoryear{De Pontieu}{2007}]{De_Pontieu2007} 
De Pontieu B., et al.: 2007, \textit{Science}, {\bf 318}, 1574. 
\href{http://adsabs.harvard.edu/abs/2007Sci...318.1574D}{ADS}. \href{http://dx.doi.org/10.1126/science.1151747}{DOI}. 

\bibitem[\protect\citeauthoryear{Dwivedi and Srivastava}{2006}]{Dwivedi2006}
Dwivedi B.N., Srivastava, A.K.: 2006, \textit{Sol. Phys.}, {\bf 237}, 1. 
\href{http://adsabs.harvard.edu/abs/2006SoPh..237..143D}{ADS}. \href{http://dx.doi.org/10.1007/s11207-006-0141-2}{DOI}. 

\bibitem[\protect\citeauthoryear{Dwivedi and Srivastava}{2010}]{Dwivedi2010} 
Dwivedi B.N., Srivastava A.K.: 2010, \textit{Curr. Sci. India}, {\bf 98}, 295. 
\href{http://adsabs.harvard.edu/abs/2010CSci...98..295D}{ADS}.

\bibitem[\protect\citeauthoryear{Hollweg}{1981}]{Hollweg1981}
Hollweg J.V.: 1981, \textit{Sol. Phys.}, {\bf 70}, 1.
\href{http://adsabs.harvard.edu/abs/1981SoPh...70...25H}{ADS}. \href{http://dx.doi.org/10.1007/BF00154391}{DOI}. 

\bibitem[\protect\citeauthoryear{Hollweg}{1990}]{Hollweg1990} 
Hollweg J.V.: 1990, \textit{Geoph. Monograph Series}, AGU, 23.
\href{http://adsabs.harvard.edu/abs/1990GMS....58...23H}{ADS}. \href{http://dx.doi.org/10.1029/GM058p0023}{DOI}. 

\bibitem[\protect\citeauthoryear{Hollweg}{1992}]{Hollweg1992}
Hollweg J.V.: 1992, \textit{Astrophys. J.}, {\bf 389}, 731. 
\href{http://adsabs.harvard.edu/abs/1992ApJ...389..731H}{ADS}. \href{http://dx.doi.org/10.1086/171246}{DOI}. 


\bibitem[\protect\citeauthoryear{Hollweg and Isenberg}{2007}]{HollwegIsenberg2007}
Hollweg J.V., Isenberg, P.A.: 2007, \textit{J. Geophys. Res.}, {\bf 112}, A08102. 
\href{http://adsabs.harvard.edu/abs/2007JGRA..112.8102H}{ADS}. \href{http://dx.doi.org/10.1029/2007JA012253}{DOI}. 


\bibitem[\protect\citeauthoryear{Hollweg, Jackson and Galloway}{1982}]{Hollweg1982}
Hollweg J.V., Jackson S., Galloway D.: 1982, \textit{Sol. Phys.}, {\bf 75}, 35. 
\href{http://adsabs.harvard.edu/abs/1982SoPh...75...35H}{ADS}. \href{http://dx.doi.org/10.1007/BF00153458}{DOI}. 


\bibitem[\protect\citeauthoryear{Jel{\'{\i}}nek et al.}{2015}]{Jelnek2015}
Jel{\'{\i}}nek P., Srivastava A.K., Murawski K., Kayshap P., Dwivedi B.N.: 2015, \textit{Astron. Astrophys.}, {\bf 581}, A131. 
\href{http://adsabs.harvard.edu/abs/2015A\%26A...581A.131J}{ADS}. \href{http://dx.doi.org/10.1051/0004-6361/201424234}{DOI}. 


\bibitem[\protect\citeauthoryear{Jess}{2009}]{Jess2009}
Jess D.B., Mathioudakis M., Erd\'elyi R., Crokett P.J., Keenan F.P., Christian D.J.: 2009, \textit{Science}, {\bf 323}, 1582. 
\href{http://adsabs.harvard.edu/abs/2009Sci...323.1582J}{ADS}. \href{http://dx.doi.org/10.1126/science.1168680}{DOI}.

\bibitem[\protect\citeauthoryear{Kudoh and Shibata}{1999}]{KudohShibata1999}
Kudoh T., Shibata K.: 1999, \textit{Astrophys. J.}, {\bf 514}, 493.
\href{http://adsabs.harvard.edu/abs/1999ApJ...514..493K}{ADS}. \href{http://dx.doi.org/10.1086/306930}{DOI}.

\bibitem[\protect\citeauthoryear{Low}{1985}]{Low1985}
Low B.C.: 1985, \textit{Astrophys. J.}, {\bf 293}, 31.
\href{http://adsabs.harvard.edu/abs/1985ApJ...293...31L}{ADS}. \href{http://dx.doi.org/10.1086/163211}{DOI}.

\bibitem[\protect\citeauthoryear{McIntosh}{1999}]{McIntosh2011}
McIntosh S.W., et al.: 2011, \textit{Nature}, {\bf 475}, 477. 
\href{http://adsabs.harvard.edu/abs/2011Natur.475..477M}{ADS}. \href{http://dx.doi.org/10.1038/nature10235}{DOI}.

\bibitem[\protect\citeauthoryear{Murawski and Musielak}{2010}]{Murawski2010}
Murawski K., Musielak Z.E.: 2010, \textit{Astron. Astrophys.}, {\bf 581}, A37. 
\href{http://adsabs.harvard.edu/abs/2010A\%26A...518A..37M}{ADS}. \href{http://dx.doi.org/10.1051/0004-6361/201014394}{DOI}.

\bibitem[\protect\citeauthoryear{Murawski et al.}{2015}]{Murawski2015}
Murawski K., Solov'ev A., Kra{\'{s}}kiewicz J., Srivastava A.K.: 2015, \textit{Astron. Astrophys.}, {\bf 576}, A22. 
\href{http://adsabs.harvard.edu/abs/2015SoPh..290.1909M}{ADS}. \href{http://dx.doi.org/10.1007/s11207-015-0740-x}{DOI}.

\bibitem[\protect\citeauthoryear{Murawski et al.}{2015}]{Murawski2015shells}
Murawski K., Srivastava A.K., Musielak Z.E., Dwivedi B.N.: 2015, \textit{Astrophys. J.}, {\bf 808}, 1.
\href{http://adsabs.harvard.edu/abs/2015ApJ...808....5M}{ADS}. \href{http://dx.doi.org/10.1088/0004-637X/808/1/5}{DOI}.

\bibitem[\protect\citeauthoryear{Murawski}{2012}]{Murawski2012} 
Murawski K., Srivastava A.K., Chmielewski P., Musielak Z.E.: 2012, 39th COSPAR Scientific Assembly, {\bf 39}, 1302. 
\href{http://adsabs.harvard.edu/abs/2012cosp...39.1302M}{ADS}.

\bibitem[\protect\citeauthoryear{Murawski}{2014}]{Murawski2014} 
Murawski K., Srivastava A.K., Musielak Z.E.: 2014, \textit{Astrophys. J.}, {\bf 788}, 8.
\href{http://adsabs.harvard.edu/abs/2014ApJ...788....8M}{ADS}. \href{http://dx.doi.org/10.1088/0004-637X/788/1/8}{DOI}.

\bibitem[\protect\citeauthoryear{Musielak}{1992}]{Musielak1992} 
Musielak Z.E., Fontenla J.M., Moore R.L.: 1992, \textit{Phys. Fluids}, {\bf 4}, 13. 
\href{http://adsabs.harvard.edu/abs/1992PhFlB...4...13M}{ADS}. \href{http://dx.doi.org/10.1063/1.860452}{DOI}.

\bibitem[\protect\citeauthoryear{Musielak and Moore}{1995}]{MusielakMoore1995} 
Musielak Z.E., Moore R.J.: 1995, \textit{Astrophys. J.}, {\bf 452}, 434. 
\href{http://adsabs.harvard.edu/abs/1995ApJ...452..434M}{ADS}. \href{http://dx.doi.org/10.1086/176314}{DOI}.

\bibitem[\protect\citeauthoryear{Musielak}{2006}]{Musielak2006} 
Musielak Z.E., Musielak D.E., Mobashi H.: 2006, \textit{Phys. Rev. E}, {\bf 73}, 036612-1-10. 
\href{10.1103/PhysRevE.73.036612}{DOI}. 

\bibitem[\protect\citeauthoryear{Musielak, Routh and Hammer}{2007}]{Musielak2007} 
Musielak Z.E., Routh S., Hammer R.: 2007, \textit{Astrophys. J.}, {\bf 659}, 650.
\href{http://adsabs.harvard.edu/abs/2007ApJ...659..650M}{ADS}. \href{http://dx.doi.org/10.1086/512776}{DOI}.

\bibitem[\protect\citeauthoryear{Mignone et al.}{2007}]{Mignone2007}
Mignone A., Bodo G., Massaglia S., Matsakos T., Tesileanu O., Zanni C., Ferrari A.: 2007, \textit{Astrophys. J. Suppl. S.}, {\bf 170}, 1. 
\href{http://adsabs.harvard.edu/abs/2007ApJS..170..228M}{ADS}. \href{http://dx.doi.org/10.1086/513316}{DOI}.

\bibitem[\protect\citeauthoryear{Mignone et al.}{2011}]{Mignone2011}
Mignone A., Zanni C., Tzeferacos P., van Straalen B., Colella P., Bodo G.: 2011, \textit{Astrophys. J. Suppl. S.}, {\bf 198}, 1.
\href{http://adsabs.harvard.edu/abs/2012ApJS..198....7M}{ADS}. \href{http://dx.doi.org/10.1088/0067-0049/198/1/7}{DOI}.

\bibitem[\protect\citeauthoryear{Nakariakov and Verwichte}{2005}]{Nakariakov2005}
Nakariakov V.M., Verwichte E.: 2005, \textit{Living Rev. Sol. Phys.}, {\bf 2}, 3. 
\href{http://adsabs.harvard.edu/abs/2005LRSP....2....3N}{ADS}. \href{http://dx.doi.org/10.12942/lrsp-2005-3}{DOI}.

\bibitem[\protect\citeauthoryear{Ofman and Aschwanden}{2002}]{Ofman2002} 
Ofman L., Aschwanden M,J.: 2002, \textit{Astrophys. J.}, {\bf 576}, L153. 
\href{http://adsabs.harvard.edu/abs/2002ApJ...576L.153O}{ADS}. \href{http://dx.doi.org/10.1086/343886}{DOI}.

\bibitem[\protect\citeauthoryear{Ofman and Davila}{1995}]{Ofman1995}
Ofman L., Davila J. M.: 1995, \textit{J. Geophys. Res-Space}, {\bf 100}, A12. 
\href{http://adsabs.harvard.edu/abs/1995JGR...10023413O}{ADS}. \href{http://dx.doi.org/10.1029/95JA02222}{DOI}.

\bibitem[\protect\citeauthoryear{Okamoto}{2007}]{Okamoto2007} 
Okamoto T.J. et al.: 2007, \textit{Science}, {\bf 318}, 1577.
\href{http://adsabs.harvard.edu/abs/2007Sci...318.1577O}{ADS}. \href{http://dx.doi.org/10.1126/science.1145447}{DOI}.

\bibitem[\protect\citeauthoryear{Okamoto and De Pontieu}{2011}]{Okamoto2011}  
Okamoto T.J., de Pontieu B.: 2011, \textit{Astrophys. J.}, {\bf 736}, L24. 
\href{http://adsabs.harvard.edu/abs/2011ApJ...736L..24O}{ADS}. \href{http://dx.doi.org/10.1088/2041-8205/736/2/L24}{DOI}.

\bibitem[\protect\citeauthoryear{O'Shea, Banerjee and Doyle}{2005}]{OShea2005}
O'Shea E., Banerjee D., Doyle J.G.: 2005, \textit{Astron. Astrophys.}, {\bf 436}, 2. 
\href{http://adsabs.harvard.edu/abs/2005A\%26A...436L..35O}{ADS}. \href{http://dx.doi.org/10.1051/0004-6361:200500120}{DOI}.

\bibitem[\protect\citeauthoryear{Perera, Musielak and Murawski}{2015}]{Perera2015}
Perera H.K., Musielak Z.E., Murawski K.: 2015, \textit{Mon. Not. R. Astron. Soc.}, {\bf 450}, 3. 
\href{http://adsabs.harvard.edu/abs/2015MNRAS.450.3169P}{ADS}. \href{http://dx.doi.org/10.1093/mnras/stv859}{DOI}.

\bibitem[\protect\citeauthoryear{Routh, Musielak and Hammer}{2010}]{Routh2010} 
Routh S., Musielak Z.E., Hammer R.: 2010, \textit{Astrophys. J.}, {\bf 709}, 1297. 
\href{http://adsabs.harvard.edu/abs/2010ApJ...709.1297R}{ADS}. \href{http://dx.doi.org/10.1088/0004-637X/709/2/1297}{DOI}.

\bibitem[\protect\citeauthoryear{Suzuki and Inutsuka}{2005}]{Suzuki2005}
Suzuki T.K., Inutsuka S.: 2005, \textit{Astrophys. J. Lett.}, {\bf 632}, 1. 
\href{http://adsabs.harvard.edu/abs/2005ApJ...632L..49S}{ADS}. \href{http://dx.doi.org/10.1086/497536}{DOI}.

\bibitem[\protect\citeauthoryear{Toro}{2009}]{Toro2009}
Toro E.: 2009, \textit{Riemann Solvers and Numerical Methods for Fluid Dynamics}, Springer-Verlag, Berlin. 
\href{http://doi.org/10.1007\%2Fb79761}{DOI}.

\bibitem[\protect\citeauthoryear{Ulmschneider, Z\"ahringer and Musielak}{1991}]{Ulmschneider1991}  
Ulmschneider P., Z\"ahringer K., Musielak Z.E.: 1991, \textit{Astron. Astrophys.}, {\bf 241}, 625.
\href{http://adsabs.harvard.edu/abs/1991A\%26A...241..625U}{ADS}. 


\bibitem[\protect\citeauthoryear{Van Doorsselaere, Nakariakov and Verwichte}{2008}]{Van_Doorsselaere2008} 
Van Doorsselaere T., Nakariakov V.N., Verwichte E.: 2008, \textit{Astrophys. J.}, {\bf 676}, L73. 
\href{http://adsabs.harvard.edu/abs/2008ApJ...676L..73V}{ADS}. \href{http://dx.doi.org/10.1086/587029}{DOI}.

\bibitem[\protect\citeauthoryear{Webb et al.}{2012}]{Webb2012} 
Webb G.M., McKenzie J.F., Hu Q., Zank G.P.: 2012, \textit{J. Geophys. Res-Space}, {\bf 117}, 5229. 
\href{http://adsabs.harvard.edu/abs/2012JGRA..117.5229W}{ADS}. \href{http://dx.doi.org/10.1029/2012JA017561}{DOI}.

\bibitem[\protect\citeauthoryear{Zhugzhda and Locans}{1982}]{Zhugzhda1982} 
Zhugzhda I.D., Locans V.: 1982, \textit{Sol. Phys.}, {\bf 76}, 77.
\href{http://adsabs.harvard.edu/abs/1982SoPh...76...77Z}{ADS}. \href{http://dx.doi.org/10.1007/BF00214131}{DOI}.

\end{thebibliography}
\end{document}